\documentclass[journal]{IEEEtran}

\usepackage{amsthm}
\usepackage{amsmath} %
\usepackage{mathtools}
\usepackage{graphicx}
\usepackage{psfrag}
\usepackage{epsf,epsfig}
\usepackage{epstopdf}
\usepackage{subfigure}
\usepackage{cite,color}
\usepackage{cite}
\usepackage{amsmath,amssymb,amsfonts}
\usepackage{algorithmic}
\usepackage{graphicx}
\usepackage{textcomp}
\usepackage{xcolor}
\usepackage{verbatim}

\DeclareMathOperator*{\argmin}{arg\,min}
\DeclareMathOperator*{\argmax}{arg\,max}

\definecolor{red}{rgb}{1,0,0}  

\newcommand{\expect}[1]{\mathbb{E}\left[#1\right]}   

\newcommand{\snr}{\mathsf{snr} } 

\newcommand{\SINR}{\mathsf{SINR} }

\newcommand{\Ccal}{\mathcal{C}}

\newcommand{\Ebb}{\mathbb{E}}

\newcommand{\Ne}{N_{\rm e}}
\newcommand{\NePAPR}{\Ne^{({\rm  min \, PAPR})}}
\newcommand{\NeCapa}{\Ne^{({\rm max \, Capa })}}
\newcommand{\NeAWGN}{\Ne^{({\rm max \, AWGN })}}
\newcommand{\Nsc}{N_{\rm sc}}
\newcommand{\Ndata}{N_{\rm data}} 
\newcommand{\Nfft}{N_{\rm fft}}
\newcommand{\Ncp}{N_{\rm cp}}

\newcommand{\jrm}{{\rm j}}

\newtheorem{Lem}{Lemma}
\newtheorem{Rem}{Remark}

\begin{document}

\title{Optimum Spectrum Extension \\for PAPR Reduction of DFT-s-OFDM}
\author{
\IEEEauthorblockN{Renaud-Alexandre Pitaval, Fredrik Berggren and Branislav M. Popovi{\'c}}
\thanks{
Copyright (c) 2026 IEEE. Personal use of this material is permitted. However, permission to use this material for any other purposes must be obtained from the IEEE by sending a request to pubs-permissions@ieee.org. 

The authors are with Huawei Technologies Sweden AB, SE-164 94 Kista, Sweden (e-mail: \{renaud.alexandre.pitaval, fredrik.b, branislav.popovic \}@huawei.com).}
}

\maketitle

\begin{abstract} 
Uplink coverage in cellular networks is constrained by the maximum UE transmit power, making peak-to-average power ratio (PAPR) reduction essential. While DFT-s-OFDM with frequency-domain spectral shaping (FDSS) achieves significantly lower PAPR than OFDM, especially with $\pi/2$-BPSK, the PAPR remains too high for higher-rate transmission. Spectrum extension (SE) combined with FDSS (FDSS-SE) can further reduce the PAPR for higher-order QAM.  This paper considers FDSS-SE with parametrized FDSS windows spanning a range of possible power ripples, as well as  arbitrary circular shifts of the subcarrier coefficients.  We optimize both the frequency shift and the SE size, and show that there exists an optimal SE size for reducing the PAPR and another one for increasing the rate. Analysis and simulations reveal that both optima largely depend on the window attenuation but are nearly invariant in proportion to the bandwidth. While the PAPR-optimal SE size is nearly invariant to the constellation order of regular QAM,  the rate-optimal SE size depends also on the SNR. 
These insights provide  practical guidelines for  
beyond-5G uplink coverage enhancement, highlighting that SE size should be individually configured according to the user's FDSS window and link quality.   
\end{abstract}


\section{Introduction}
Uplink transmission is typically the coverage bottleneck of modern wireless cellular systems due to the limited transmission power of user equipment (UE). As a result, uplink coverage enhancement remains an on-going topic in 3GPP standardization, as it has been for 5G~\cite{ReportRAN1_102,RP-251862} and expected to continue into 6G~\cite{RP-251881}. 

In coverage-limited scenarios, transmitting at the highest possible average power is desirable to maximize the link quality. 
However, the dynamic range of the transmitted signals, commonly measured by its peak-to-average power ratio (PAPR), often forces the UE to back off from its maximum power to avoid distortion from  peaks penetrating the non-linear region of the power amplifier. This back-off prevents uncontrolled distortion but directly reduces the signal-to-noise ratio (SNR).

Orthogonal frequency division multiplexing (OFDM) is the dominant modulating waveform of wireless communications and forms the basis of 3GPP standards for both uplink and downlink. Discrete Fourier transform-spread OFDM (DFT-s-OFDM) 
is an uplink option in 5G NR that reduces signal fluctuations compared to conventional OFDM.
In DFT-s-OFDM, data symbols undergo DFT precoding before OFDM modulation, 
producing a single-carrier-like waveform since each modulation symbol’s energy is distributed across all allocated subcarriers. 
Reciprocally, in the time domain, a large fraction of a symbol’s energy is concentrated in a short interval, leading to a pulse-multiplexing of the data.  
While DFT-s-OFDM improves the PAPR over OFDM, the PAPR remains high with regular quadrature amplitude modulation (QAM), motivating the need for further low-PAPR techniques in both research literature and 3GPP standardization.
One approach is to modify the symbol constellation, such as employing $\pi/2$-BPSK (supported in 5G uplink in addition to legacy 4G constellations), or other rotated-QAM schemes~\cite{KimTVT18,TENCON18}.   
Another add-on feature is to  apply a spectral window, referred to as  frequency-domain spectral (FDSS),  reducing subsequently the PAPR at the cost of controlled self-interference.  This is implicitly supported in 5G -- for $\pi/2$-BPSK from its first release via relaxed spectrum error vector magnitude (EVM) mask requirements for proprietary signal implementations~\cite{3GPPTS38.101-2},   and recently extended to QPSK in 5G-Advanced Rel. 18  by allowing power boosting --   
enabling FDSS as a transparent PAPR-reduction method without specifying an exact window.
  
Spectrum extension (SE) is another complementary PAPR-reduction technique, 
in which the DFT-spread data sequence is periodically extended -- typically before applying an FDSS window. SE has been discussed in 4G~\cite{R1-050702,Kawamura06}, considered for 5G~\cite{RP-213579,Nokia21}, and now proposed again for 6G~\cite{SamsungSPAWC2024,6GWS-250004,6GWS-250036}.  The motivation is to enable low-PAPR also for higher-order modulation such as QPSK (i.e. 4-QAM), thereby improving coverage at higher-data rates than those supported by $\pi/2$-BPSK currently in 5G.

FDSS with SE (FDSS-SE) has been investigated in several works, e.g.,~\cite{Kawamura06,SlimaneTVT07,KimTVT18,Nokia21}. 
Early studies~\cite{R1-050702,Kawamura06,SlimaneTVT07}, often coupled the FDSS window design to the SE size,  implicitly assuming a fixed SE size without adaptation to UE capability or other transmission parameters. These works did not examine the effect of varying SE size for a given FDSS roll-off factor. Often, academic works with a more systematic approach than 3GPP standardization, such as~\cite{KimTVT18}, optimize the FDSS window for a given system configuration, which does not align well with practical flexibility of current systems.  Moreover, \cite{R1-050702,Kawamura06} assumed tight spectrum-flatness requirements, restricting FDSS transition to narrow bands -- limiting potential benefits compared to the broader transition bands allowed by the looser spectrum masks introduced in 5G. 
SE was also implicitly assumed to reduce PAPR at the expense of spectral efficiency, and so in order to mitigate this,  preferably having extended subcarriers outside of the allocated band and overlapping with other transmissions.  

The more recent work~\cite{Nokia21}, which had a strong influence on the 5G Rel-18 study on FDSS-SE for uplink coverage enhancement, evaluated FDSS-SE at a fixed spectral efficiency, i.e. compensating the bandwidth loss by using higher-rate channel codes, and using  a double-side symmetric SE with a fixed SE size at 25\% of the bandwidth. 
QPSK with FDSS-SE was shown to outperform QPSK with FDSS but without SE, meaning that 
the power back-off reduction gain outweighed the bandwidth efficiency loss. 
In that setup, extended subcarriers did not overlap with other bands, enabling frequency-diversity combining of repeated symbols at the receiver.

While earlier academic works such as~\cite{SlimaneTVT07} considered single-side extension, symmetric SE as in~\cite{Nokia21} -- already proposed in early 3G standardization~\cite{R1-050702,Kawamura06} -- has become the de facto approach in 3GPP~\cite{SamsungSPAWC2024,6GWS-250004,6GWS-250036}. Symmetric SE corresponds to applying  a particular circular shift  to the subcarrier coefficients before single-side SE. Alternatively, in 3GPP Rel.18 study, we proposed asymmetric SE with optimized cyclic shifts~\cite{R1-2208412,R1-2300090}, derived as a by-product of the optimized constellation rotation approach in~\cite{KimTVT18,TENCON18}. Recently, similar asymmetric SE approaches have been studied in~\cite{IEICE25,IEICE25_NTT} and are proposed for 6G~\cite{R1-2506306}. Cyclic shifting is easier to implement and maintains compatibility with legacy systems, while incurring only negligible precision loss compared to optimal constellation rotation~\cite{IEICE25}.

Moreover, while it is well established that SE can enable further  PAPR reduction on top of FDSS, it is less recognized that an excessively large SE may instead increase it. The prevailing view being that SE size should be limited, but solely by bandwidth-efficiency considerations. 
Conversely, if ignoring the PAPR effect, SE is commonly regarded as being inherently detrimental to spectral efficiency. 
In contrast,  
this paper draws on new insights into FDSS-SE design, showing that for a given FDSS window there exists both an optimal SE size for PAPR minimization -- beyond which PAPR performance degrades, and an optimal SE size for rate maximization -- below which rate decreases. 
Hence, if in 6G the FDSS windows remain proprietary and UE-specific as in 5G, i.e., implemented by manufacturers and not optimizable by the network, then enforcing a single SE size for all UEs may be suboptimal and even detrimental for some of them.

During 5G Rel. 18 standardization~\cite{R1-2208412,R1-2300090}, we already pointed-out that PAPR minimization could be achieved with different SE sizes depending on the considered FDSS window. However, evaluations in 3GPP were conducted only for a few (typically competing) designs with a limited and heterogeneous  set of specific FDSS windows and configurations, so general performance trends and behavior were not well established. 
This paper refines the PAPR analysis from our earlier 3GPP  contributions with stronger theoretical support, as well as a rate analysis showing that SE  also helps mitigate the SNR loss from FDSS. 
We use a generic FDSS-SE model with arbitrary frequency-shifts, encompassing prior single-side and symmetric definitions, and consider two single-parameter FDSS window families to evaluate the performance behaviors as a function of FDSS attenuation. 
Namely, we introduce a single-parameter ``deformed Hann'' window family covering FDSS windows used in 3GPP; and we also use a Kaiser window proposed for DFT-s-OFDM in~\cite{Mauritz06}  as a well-known approximation of the prolate spheroidal wave
functions for optimal time- and band- limited localization~\cite{SlepianPollak}. 

In passing, we revisit the relevance of the PAPR and the alternative cubic-metric (CM), the latter  typically preferred in 3GPP RAN1 study but actually shown here to be non-monotonic with FDSS shaping and thus less likely to align with power-derating trends for FDSS-SE.  

We then address the PAPR minimization for FDSS-SE with a fixed FDSS window, leveraging and adapting the bounds from~\cite{KimTVT18,TENCON18} to determine the optimal SE size and frequency shift. 
We show that symmetric extension in~\cite{Nokia21} is optimal for $\pi/2$-BPSK but not for regular QAMs, where single-side extension~\cite{SlimaneTVT07} performs better but still not optimal. The optimal shifts for regular QAMs are provided and  its PAPR gain shown to be well-anticipated by the bounds.  
More importantly, we study the optimal SE size for PAPR minimization.  
This optimum depends strongly on the FDSS window, differs between $\pi/2$-BPSK and regular QAMs, and is largely independent of the SE’s frequency shift or the allocated bandwidth (in proportion). 
All together, the optimized  shift and SE size enable us to evaluate the best PAPR gain of FDSS with SE compared to FDSS without SE, showing that FDSS-SE provides the best PAPR reduction with QPSK compared to $\pi/2$-BPSK or higher-order QAMs.

Next, we examine the spectral efficiency of FDSS-SE.  
While conventional wisdom would suggest that SE always reduces rate, we derive a generic capacity expression for DFT-s-OFDM with FDSS-SE and the combining receiver of~\cite{Nokia21}.  
Our generic FDSS-SE approach simplifies the prior formulation of this receiver, and the capacity result  
shows that there is another SE size for maximizing the rate, dependent of the FDSS window and nearly invariant of the bandwidth, but, unlike the PAPR-optimal SE size, dependent also of the SNR.  

For both metrics, PAPR and rate, analysis and results are supported by a semi-analytical approach based on numerical optimization of theoretical bounds. The PAPR-optimal and rate-optimal SE sizes are in general different, providing a bounded range of relevant SE sizes within which a balanced choice can improve both metrics simultaneously.

The rest of the paper is as follows:  
Section II presents the  DFT-s-OFDM transmitter model with FDSS-SE. Section III discusses accurate PAPR computation, compare with the CM its relevance for FDSS-SE study. 
Section IV analyzes FDSS-SE optimization for PAPR minimization. Section V covers the receiver model and spectral efficiency aspects. The paper is concluded in Section VI.

\section{DFT-s-OFDM Transmission}
\subsection{DFT-s-OFDM with FDSS and Spectrum Extension}

\begin{figure}[t]
\centering
\vspace{-0.0cm} 
\includegraphics[width=.5\textwidth]{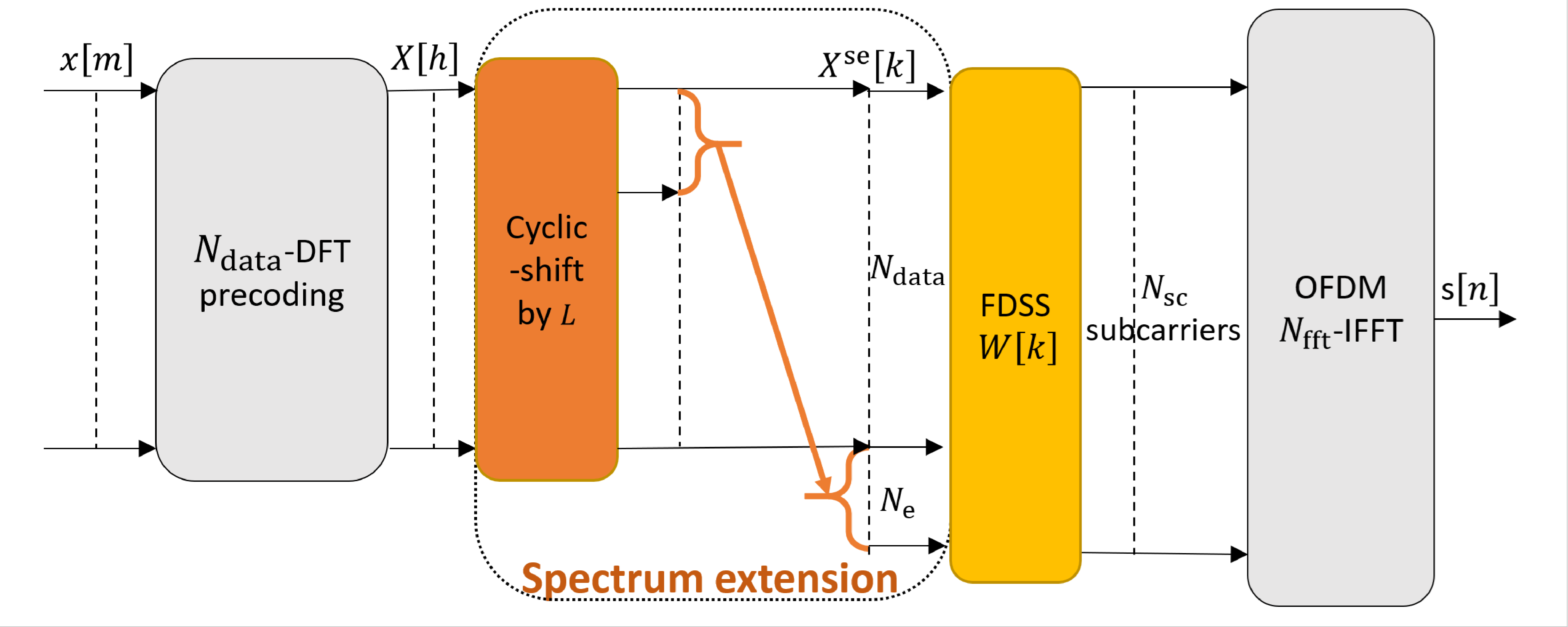}
\vspace{-0.4cm}
	\caption{Block diagram of DFT-s-OFDM with FDSS-SE. \label{fig:FDSS-SE}}
\vspace{-0.2cm}
\end{figure}

The DFT-s-OFDM transmitter with FDSS-SE is schematized in Fig.~\ref{fig:FDSS-SE}
A DFT-s-OFDM symbol with cyclic prefix (CP) is defined for samples $-\Ncp\leq n\leq \Nfft-1$ as 
\begin{equation}  \label{eq:OFDMsignal}
s[n] = \frac{1}{\sqrt{\Nfft}} \sum_{k=0}^{\Nsc-1}  W[k] X^{\rm se}[k] e^{\jrm \frac{2 \pi nk}{\Nfft} }  
\end{equation}  
where  
\begin{gather}\label{eq:Xdftse}
X^{\rm se}[k] = X[(k+L) 
\;{\rm mod}\; \Ndata]
\\ \text{for }k=0,\ldots,\Nsc-1, \nonumber 
\end{gather}
are the spectrum-extended version of the subcarrier coefficients 
\begin{gather} \label{eq:Xdft}
X[h] = \frac{1}{\sqrt{\Ndata}} \sum_{m=0}^{\Ndata-1}  x[m] 
 e^{-\jrm \frac{2 \pi hm}{\Ndata} },
\\ \text{for } h=0,\ldots,\Ndata-1, \nonumber   
\end{gather}
which are the DFT precoding  of constellation symbols $x[m]\in \Ccal$,  
where  
\begin{itemize} 
\item	$\Ndata$ is the number of modulation constellation symbols, which are assumed to be  independently and identically distributed (i.i.d.) QAM symbols with zero-mean and  unit average energy, i.e. $\Ebb[|x[m]|^2 ]=1$,
\item	$\Nsc $	is the number of modulated OFDM subcarriers, 
\item	$\Ne=\Nsc- \Ndata$ is the number of subcarriers used for SE, referred  as the SE size,
\item	$L$ is a circular shift value with $({\rm mod}\; N)$ being the modulo-$N$ operator,
\item $W[k]$ is a FDSS window which is assumed real and symmetric, and normalized as\footnote{Note that in~\cite{KimTVT18}, the FDSS window power is set to $\Ndata$. The choice of normalization here makes the transmit power constant and independent of the SE size and only dependent of the bandwidth allocation. This matches practical consideration as considered in 3GPP study on coverage enhancement.} 
\begin{equation}
\sum_{k=0}^{\Nsc-1}|W[k]|^2 = \Nsc
\end{equation}
\item $\Nfft$ is the IFFT size of the OFDM modulation and $\Ncp $ is the CP length.
\end{itemize}

\begin{figure}[t]
\centering
\vspace{-0.0cm} 
\includegraphics[width=.48\textwidth]{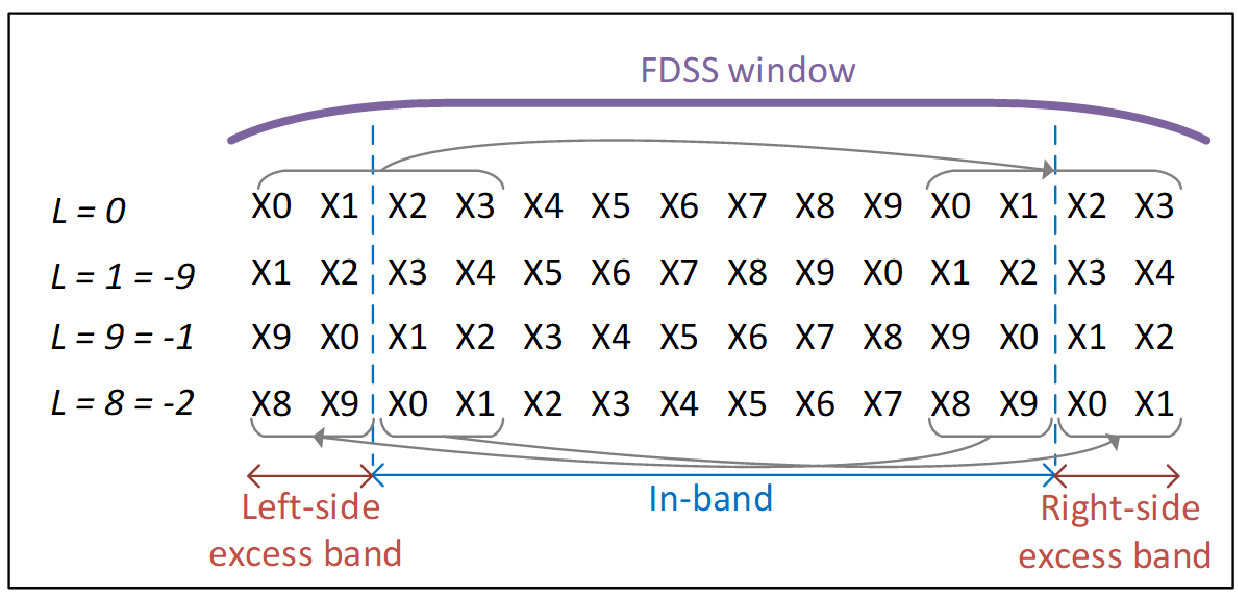}
\vspace{-0.20cm}
	\caption{Illustration of spectrum-extended data sequence as a function of the shift parameter $L$ with $\Ndata=10$ symbols and $\Ne=4$. \label{fig:SE_illustration}}
\vspace{-0.2cm}
\end{figure}

The model encompasses different SE approaches. 
For example, in~\cite{Nokia21}, the spectrum extension is defined as double-side and symmetric with $L=\Ndata-\Ne/2 =-\Ne/2 \;({\rm mod}\; \Ndata)$, while in~\cite{SlimaneTVT07,KimTVT18} extension is one-side with $L=0$.
This interpretation as symmetric/asymmetric or two-/single-side is to a large extent artificial, and can be applied to any shift value.   
Indeed, as illustrated in Fig.~\ref{fig:SE_illustration}, for any $L$, all SE definitions have the same symmetries, and can be seen as one- or two-side extensions. The complete set of symbols $\{X[0],\ldots,X[\Ndata-1]\}$ is always in the ``in-band'', as defined in~\cite{Nokia21}, up to a cyclic-shift by $L+\Ne/2$ symbols; and the left-side excess-band symbols are always the repetition of symbols of the right-side in-band edge, similarly for the right-side excess band. 
The benefit of symmetric SE as in~\cite{Nokia21} may lie only in a more convenient indexing. 

\subsection{Pulse Time-Multiplexing Interpretation}
The DFT-s-OFDM transmission can be reformulated as a time-multiplexing of $\Ndata$-pulses $\{p_m [n]\}_{m=0}^{\Ndata-1}$. By combining~\eqref{eq:Xdft}, \eqref{eq:Xdftse} and~\eqref{eq:OFDMsignal}, we obtain
\begin{equation} \label{eq:sMP}
s[n] =   \frac{1}{\sqrt{\Nfft }} \sum_{m=0}^{\Ndata-1}x[m] p_m [n]
\end{equation}
with time-domain pulses
\begin{equation} \label{eq:pm}
p_m [n] =e^{-\jrm \frac{2\pi L}{\Ndata}m}  p_0\left[n-\frac{\Nfft}{\Ndata} m \right] 
\end{equation} 
which are different time and phase shifted versions of the same kernel pulse-shaping filter 
\begin{equation} \label{eq:h}
p_0[n] =  \frac{1}{\sqrt{\Ndata }} \sum_{k=0}^{\Nsc-1} W[k]  e^{\jrm \frac{2\pi k}{\Nfft}n}.
\end{equation} 
In the case of no FDSS windowing, $W[k]=1$, $\forall k$, we recover the conventional DFT-s-OFDM pulse, in the form of a Dirichlet kernel, with $\Nsc$ modulated subcarriers: 
\begin{equation} \label{eq:p0sinc}
p_0[n]=\frac{e^{\jrm \frac{\pi}{\Nfft}  n(\Nsc-1)} }{\sqrt{ \Ndata }}   \frac{\sin \left( \pi \frac{\Nsc}{\Nfft} n\right) }{\sin \left(  \pi \frac{1}{\Nfft} n\right)},
\end{equation}
and then the pulses~\eqref{eq:pm} are orthogonal; otherwise, with FDSS, they are in general not orthogonal anymore. 

\subsection{FDSS Windows}
We consider FDSS window families with convenient single shaping parametrization that characterizes their \emph{maximum FDSS power ripple} defined  as $20 \log_{10} \frac{\min_k  W[k]}{\max_k  W[k]}$ [dB].

\emph{Deformed Hann Window:} 
We define a deformed Hann window with  shaping parameter $0\leq \beta\leq 1$,
\begin{equation} \label{eq:ModHanWin}
W_{\beta}[k] =\frac{1}{\omega}\left(1-\frac{1-\beta}{1+\beta} \cos  \left(\frac{2\pi k+\pi}{\Nsc}\right)\right)   
\end{equation}
where the normalization factor is $\omega = \sqrt{1+\frac{(1-\beta)^2}{2(1+\beta)^2}}$.  

With  $\beta  = 0  $, we recover the classical Hann window up to normalization, while with $\beta  = 1 $ it corresponds to no FDSS shaping. 
This window is derived based on 3-tap filters commonly-used by industry in 3GPP, see Appendix~\ref{Proof:3tapFilter}. In this convenient form,  the maximum power ripple of this FDSS window is tunable and directly given by its shaping parameter as $\approx 20 \log_{10}\beta$. Moreover, the FDSS windows from 3-tap filters are not exactly symmetric and we applied an half index shift to correct this, which has a noticeable effect only for very narrow  bands.  
Specifically, the specific 3-tap filter used in~\cite{R1-1901117,R1-2210880,R1-2305484} are equivalently obtained as $W_{\beta}[k-1/2]$ using~\eqref{eq:ModHanWin} with $\beta = -11$ and $-14 $ dB, respectively.

\emph{Kaiser Window:}  
The Kaiser window proposed for DFT-s-OFDM in~\cite{Mauritz06} is defined with shaping parameter $\kappa$  as 
\begin{equation} \label{eq:Kaiser}
W_{\kappa}[k] =\frac{1}{\omega I_0\left( \kappa \right)} I_0\left( \kappa \sqrt{1-\frac{(k-\gamma)^2}{\gamma^2}}\right) 
\end{equation}  
where $\gamma = \frac{\Nsc-1}{2}$ and $I_0(\cdot)$ represents the zeroth-order modified Bessel function of the first kind; and $\omega$ is the normalization coefficient. The maximum of the window is at $W_{\kappa}[\Nsc/2]\approx 1/\omega$, while its minimum is at  $W_{\kappa}[0]= \frac{1}{\omega I_0\left( \kappa \right)}$, so that its maximum power  ripple is $\approx-20 \log_{10} \left( I_0\left( \kappa \right)\right) $. 

This window family was considered by us in 3GPP Rel. 18 study~\cite{R1-2208412}. 
It is often close in shape to the truncated RRC (TRRC) design proposed in~\cite{Nokia21}, considered  also in 3GPP Rel. 18. However, the TRRC window is slightly less convenient for study as it has two parameters; and for  a given parameter set,  a shape that changes with the bandwidth size when the bandwidth is small.  

\begin{figure}[t]
\vspace{-0.2cm} 
\subfigure[Deformed Hann window]{\includegraphics[width=.5\textwidth]{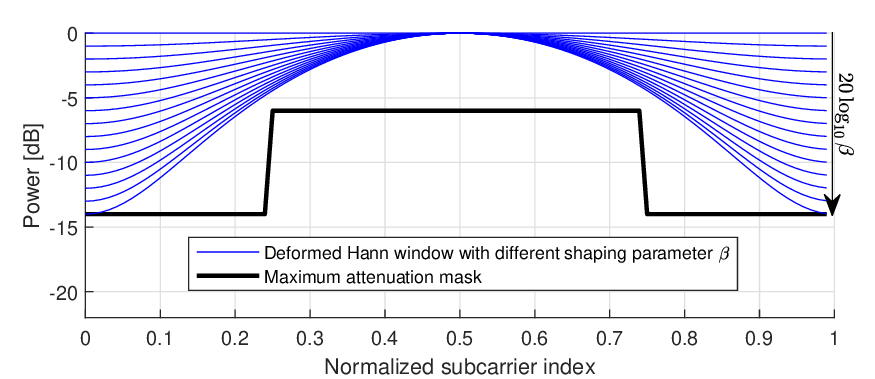}}
\subfigure[Kaiser window]{\includegraphics[width=.5\textwidth]{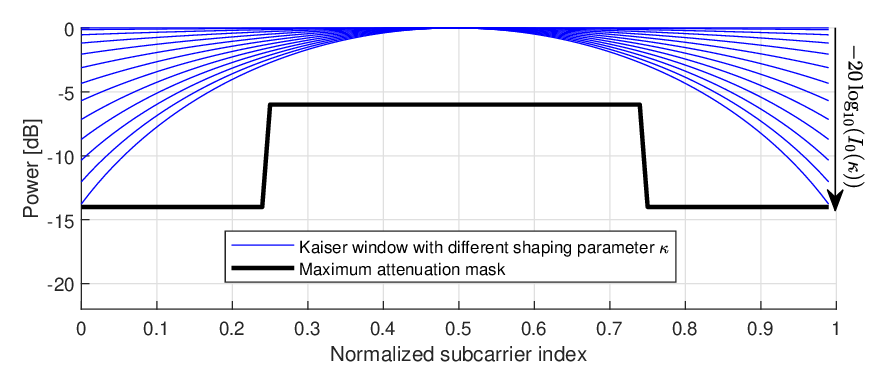}}
\vspace{-0.4cm}
	\caption{Two FDSS window types with single shaping parameters characterizing their maximum FDSS power ripple. \label{fig:HannWin}}
\vspace{-0.2cm}
\end{figure}

Fig.~\ref{fig:HannWin} shows the variation of the window shapes obtained by the Hann and Kaiser window types for different $\beta$ and $\kappa$, along the maximum FDSS attenuation permitted with $\pi/2$-BPSK according to the specification of the EVM equalizer spectral flatness requirements in~\cite{3GPPTS38.101-2}.

\section{PAPR: Definition and Relevance}

Given an OFDM symbol, we define its PAPR as
\begin{equation}  
{\rm PAPR } =  \frac{  \max_{0\leq n<\Nfft} |s[n]|^2}{\expect{|s[n]|^2}} 
\label{eq:PAPR_Exp}
\end{equation} 
where $\expect{\cdot}$ is the statistical expectation. Otherwise stated, $\Nfft= 2048$ in numerical evalutations. PAPR of the digital OFDM signal provides an accurate PAPR estimation of the corresponding transmitted analog OFDM signal when oversampling is at least $\Nfft/\Ndata = 4$~\cite{PAPROverview}. 
 
Before analyzing the PAPR of FDSS-SE, we first point out some considerations on the definition and relevance of the PAPR metric,  which appear to be	insufficiently examined in the literature.

 \subsection{Statistical Mean Power versus Instantaneous Mean Power}
The PAPR in~\eqref{eq:PAPR_Exp}  is computed using the statistical mean power (i.e., the expectation in the denominator) as in many academic works, e.g.,~\cite{SlimaneTVT07,PAPROverview,KimTVT18}.   Alternatively, several other works define the PAPR using the instantaneous arithmetic mean power over a considered signal lengh $N$, i.e. \mbox{ ${\rm PAPR }' =  \frac{ { \max_{0 \leq n < N} } |s[n]|^2}{\frac{1}{N}\sum_{n=0}^{N-1}{|s[n]|^2}}$}, and for this, often the signal length is one OFDM symbol  as, e.g., in~\cite{SamsungSPAWC2024,IEICE25_NTT,Kant,LiWei,HuangPengfei,AliAfan}.

With FDSS, the discrepancy between these two PAPR definitions becomes more pronounced, especially when the number of subcarriers and/or OFDM symbols is small. This is because higher-order constellations combined with FDSS require more statistical realizations of the constellation symbols to average out the inherent randomness.
 
While the instantaneous average power may be argued to be more meaningful since it characterizes the PAPR for a specific signal realization as experienced by hardware;  practical OFDM transmissions rarely consist of a single OFDM symbol, and the overall signal length can vary. Therefore, for a generic and analytically consistent evaluation based on a single OFDM symbol, the statistical mean power definition is preferable, if possible.

\begin{figure*}
  \begin{minipage}{0.63\textwidth}
\centering
\vspace{-0.2cm} 
\subfigure[$\pi/2$-BPSK \label{fig:PAPRvsCMbpsk}]{\includegraphics[width=5.6cm]{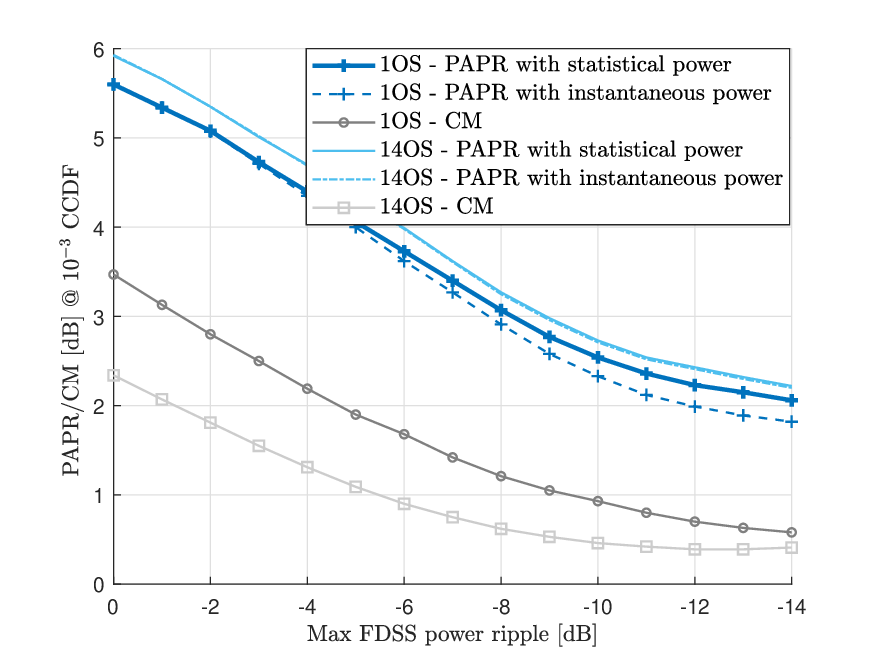}}
\subfigure[QPSK]{\includegraphics[width=5.6cm]{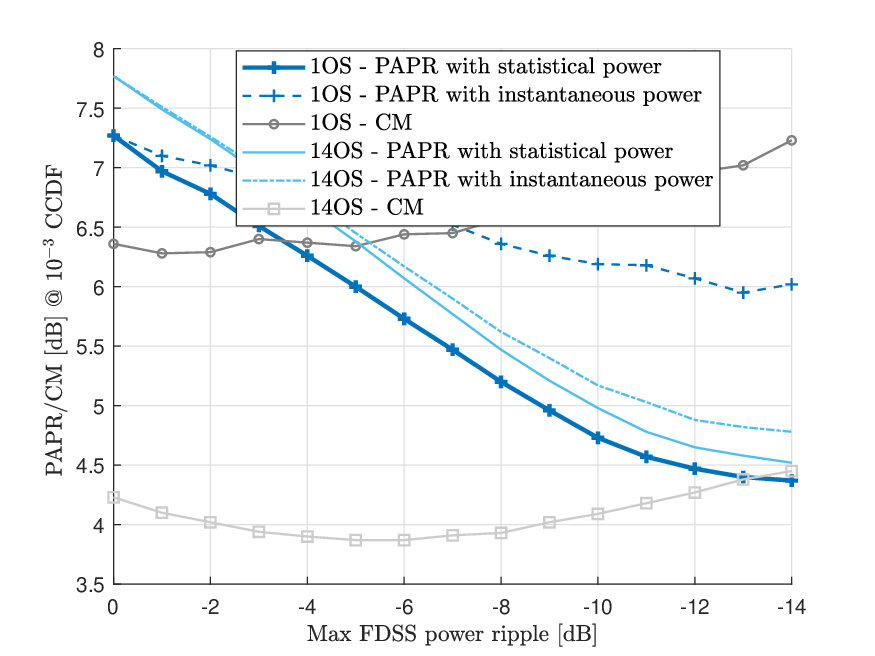}}
\vspace{-0.2cm}
	\caption{Inconsistencies among PAPR definitions when considering FDSS and QPSK in narrowband. \label{fig:PAPRvsCM}}
\end{minipage}
\hspace{0.1cm}
\begin{minipage}{0.33\textwidth}
\centering
\vspace{-0.2cm} 
\includegraphics[width=5.5cm]{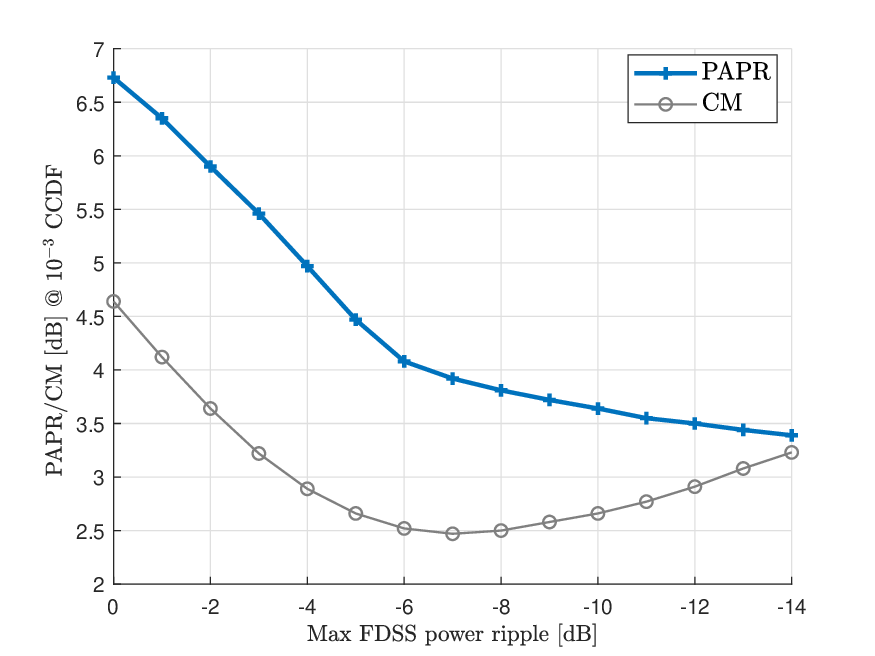}
\vspace{0.0cm}
	\caption{Inconsistent behavior of CM and PAPR for  FDSS-SE with QPSK.
	\label{fig:CMvsPAPRwSE}}
\end{minipage}
\vspace{-0.4cm}
\end{figure*}

Fig.~\ref{fig:PAPRvsCM} illustrates these considerations with $\Nsc =24$ and deformed Hann window,  where the $10^{-3}$ complementary cumulative distribution function (CCDF) level of the PAPR is plotted along increasing FDSS shaping attenuation. The PAPR is computed for both a single OFDM symbol (1OS) and a 14 CP-OFDM symbol signal\footnote{The maximum in the numerator of \eqref{eq:PAPR_Exp} is applied on the corresponding signal, i.e. for 14 OFDM symbols the peak is taken over a larger sample set, and so its PAPR is larger than with one OFDM symbol.} (14OS).  
In the absence of FDSS (0 dB),  both PAPR definitions yield the same value for  $\pi/2$-BPSK and QPSK. However, as  FDSS 
shaping increases, the two PAPR definitions diverge. 
This difference is small for $\pi/2$-BPSK, and for QPSK when signals comprise 14 CP-OFDM symbols.  
Similarly, for a single OFDM symbol with a larger number of subcarriers, the PAPR based on instantaneous mean power would approach that based on statistical mean power.  
However, as shown, for the case of QPSK with FDSS, the PAPR computed using the instantaneous mean power  is  significantly  higher  for single narrowband OFDM symbol than other cases, and thus fails to consistently reflect performance trends.

\subsection{Linear Fitting of Power De-rating by PAPR and CM}
The PAPR quantifies the envelope fluctuation of a signal. Large envelope peaks either drive the power amplifier into its nonlinear region or require transmit power back-off to avoid distortion, which directly reduces the link SNR. 
In 3GPP systems, this back-off is constrained by a maximum power reduction (MPR), defined to ensure compliance with RF requirements such as distortion and adjacent-channel interference.
In practice, the required MPR depends on the allocated bandwidth size and its position within the channel band, with larger allocations or edge allocations typically requiring larger MPR margins. Higher-order modulations generally exhibit higher PAPR and therefore incur larger MPR.

%
The cubic metric (CM) has been introduced  during late 3G UMTS standardization as an alternative to PAPR for estimating power de-rating, motivated by nonlinear amplifier modeling~\cite{R1-060023}. 
From its inception, CM has always been used together with linear regression coefficients fitted to measurement data. 
 In fact, the 3G UMTS specification~\cite{3GPPTS25.101} includes multiple shifted, scaled, and quantized versions of CM depending on the system configuration. 
For 4G LTE, the CM fitting was updated~\cite{R1-040642}  based on OFDM signal measurements, and different regression coefficients were recommended for two different bandwidth configurations.  
Since then, 5G NR has reused these definitions from~\cite{R1-060023}, originally derived for 4G, and in some cases from 3G work, see e.g.,~\cite{R1-163437}. However, applying the early-4G linear fit often leads to negative power de-rating estimates when evaluating modern low-PAPR waveforms, such as DFT-s-OFDM with $\pi/2$-BPSK and FDSS as well as other schemes, see e.g.~\cite{Qin2022}, which raises questions about the practical meaning of these CM-based estimates.

It is worth emphasizing that CM may provide a much better power de-rating estimation than PAPR, primarily because it is always used with fitted regression coefficients from measurements, whereas PAPR is typically evaluated in its raw form, without linear fitting.  
However, nothing prevents applying a linear regression to PAPR as well. 
In fact, the 3GPP contribution introducing CM~\cite{R1-060023} also considered a linear fitting of PAPR, $\approx 0.855({\rm PAPR} -3)$, showing that power de-rating correlates reasonably well with PAPR after regression,  although CM provided a slightly better overall fit. 
 More recent evaluations, such as~\cite{ICCT17}, confirms  this trend.


In any case,  a single-regression of neither CM nor PAPR could fully predict realistic power de-rating in all scenarios. 
In particular, 
both metrics in their raw forms fail to capture the impact of bandwidth size and allocation position~\cite{R1-2305484}.  

\subsection{Non-Monotonic CM Behavior with FDSS}

In 3GPP Rel-18 study on coverage enhancement, CM-based power de-rating estimation 
was  considered by  RAN1 as an intermediate step prior to RAN4 MPR evaluations~\cite{R1-2208412}. This CM-based approach initially influenced the FDSS-SE designs in RAN1, but its reliability was later challenged, and many RAN1 participants eventually relied on direct MPR evaluations~\cite{R1-2300090, R1-2305484}.

We point-out a malfunctioning  behavior of CM when applied with FDSS and higher-order QAM modulations: as FDSS shaping increases, CM follows a bell-shaped curve. As shown in Fig.~\ref{fig:PAPRvsCM}, unlike PAPR,  which improves systematically with larger FDSS attenuation,  CM exhibits this trend only for $\pi/2$-BPSK, but not for QPSK. This inconsistency  between PAPR and CM is further observed in Fig.~\ref{fig:CMvsPAPRwSE} for FDSS-SE and a larger bandwidth ($\Nsc = 96 $, $\Ne = 10$), where the bell-shape of CM becomes even more pronounced.   

This behavior does not align with an expected power de-rating prediction. Intuitively, as the signal spectrum becomes more confined, adjacent channel interference decreases, which is one of the main conformance issues limiting the transmit power. In fact, evaluations from 3GPP Rel-18 study~\cite{R1-2305484,R1-2300090} confirms that power de-rating gains are larger with more aggressive windows. 
Moreover, it is unlikely that a more confined spectrum would lead to power de-rating gain with $\pi/2$-BPSK but a loss with QPSK. 

Realistic MPR evaluations  are beyond the scope of this paper, but this suggests that  the presumed superiority of CM over PAPR for power de-rating prediction should be reconsidered in the context of FDSS-SE. Prior works such as~\cite{ICCT17}, which showed better prediction with  CM or other metrics, did not account for the impact of FDSS. Therefore, in this paper, we retain the classical PAPR metric as, not necessarily the most accurate, but a consistent and analytically tractable indicator of power de-rating performance trends.

\section{Spectrum Extension Minimizing the PAPR}

In this section, we study the PAPR of FDSS-SE as a function of the SE size $\Ne= \Nsc-\Ndata $ and the frequency-shift $L$,
assuming that  $\Nsc$ and the FDSS window are pre-configured and fixed.

\subsection{PAPR Bounds}
We begin by expressing PAPR bounds that will help to characterize and verify the PAPR behavior.  Based on the pulse multiplexing interpretation~\eqref{eq:sMP} with pulses $p_m[n]$ defined in~\eqref{eq:pm}, a simple adaptation and generalization of the approach in~\cite{KimTVT18} and~\cite{TENCON18} (see Appendix~\ref{App:PAPRbounds})  yields the following PAPR upper bound, which  depends only on $\Ne$, $L$, and the symbol constellation. 
\begin{Lem}
For an SE size $\Ne$ with shift $L$ and constellation $\Ccal$ with largest amplitude $A_{\Ccal}$, we have 
\begin{equation}
{\rm PAPR } \leq   {\rm PAPR }^{\rm U}(\Ccal,\Ne,L) \leq {\rm PAPR }^{\rm GU}(A_{\Ccal},\Ne) 
 \end{equation}
with 
\begin{equation} \label{eq:PAPRub}
 \!\!\!\! {\rm PAPR }^{\rm U}(\Ccal,\Ne,L) =  \frac{A_{\Ccal}^2 }{\Nsc}  \max_n  \left\{ \sum_{i,j= 0}^{\Ndata-1} |p_i[n]||p_j[n]| u_{i,j}^{\Ccal}   \right\}
\end{equation} 
where $p_i[n]$ are the time-domain pulses~\eqref{eq:pm} and 
\begin{equation} \label{eq:uij}
u_{i,j}^{\Ccal} = \max_{\omega \in \Omega_{\Ccal}} \left| \cos \left((i-j)\left[
\phi - \frac{(2L+\Ne-1)}{\Ndata} \pi \right] +\omega \right)\right|,
\end{equation}
$\Omega^{\Ccal}$ being the set of phase differences among constellation symbol modulo $\pi$. 

This is further upper-bounded by the more general bound  
\begin{equation} \label{eq:PAPRub2}
 {\rm PAPR }^{\rm GU}(A_{\Ccal},\Ne)  = 
\frac{ A_{\Ccal}^2 }{\Nsc}  \max_n  \left\{ \sum_{i,j= 0}^{\Ndata-1} |p_i[n]||p_j[n]| \right\} .
\end{equation}
\end{Lem}

More precisely: 
\begin{itemize} 
\item For $\pi/2$-BPSK,  $\phi = \frac{\pi}{2}$, $A_{\Ccal} = 1$, and $\Omega_{\Ccal} = \{ 0 \}$. 
\item For QPSK,  $\phi = 0$, $ A_{\Ccal} = 1$, and $\Omega^{\Ccal} = \{ 0, \frac{\pi}{2}\}$. 
\item For 16-QAM,  $\phi = 0$, $ A_{\Ccal}^2= 1.8$, and $\Omega^{\Ccal} $ contains 10 different values. 
\end{itemize} 

As the constellation grows, the set  of angles $\Omega^{\Ccal} $ becomes a denser sampling of $[0,\, \pi]$ so the first upper bound converges to the general upper bound ${\rm PAPR }^{\rm U}(\Ccal,\Ne,L) \to {\rm PAPR }^{\rm GU}(A_{\Ccal},\Ne)$. 

Moreover, for regular QAM, note that the maximum symbol power $A_{\Ccal}^2$ occurs only for a subset of constellation points at the four edges, forming a QPSK sub-constellation with larger amplitude. We can therefore expect that symbol combination of this sub-QPSK constellation will yield some of the highest signal peaks, leading to: 
\begin{equation} \label{eq:PAPRubQPSK}
{\rm PAPR }^{\rm U}(\Ccal_{\rm QAM},\Ne,L) \approx  A_{\Ccal_{\rm QAM}}^2 \times {\rm PAPR }^{\rm U}(\Ccal_{\rm QPSK},\Ne,L).
\end{equation}  
	
To compute these bounds, the maximum over one OFDM symbol duration $0\leq n \leq \Nfft-1$ must be found. The computation complexity of this step can be greatly reduced by noting that the function inside the max-operator is periodic, since all pulses in~\eqref{eq:pm} are regular time-shifted versions of the same kernel. It is therefore sufficient to  search over $\Nfft/\Ndata$ consecutive samples to find the maximum.


\subsection{Best Shifting Value}
The best shifting value $L$ for FDSS-SE follows from the phase difference between two neighboring pulses with indices $m$ and $(m+1)$  (see Appendix~\ref{App:EqPhase}):
\begin{equation}
\angle \frac{p_{m+1}[n]}{p_{m}[n]}= - \frac{2\pi}{\Ndata}\left(L +\frac{(\Ne-1)}{2} \right) \; ({\rm mod}\,  \pi), \label{eq:pulsePhdiff}
\end{equation}  
which corresponds to the term~\eqref{eq:uij} inside the bound~\eqref{eq:PAPRub}.

\emph{For $\pi/2$-BPSK}, the purpose of the $\pi/2$-rotation is to ensure that BPSK symbol of neighboring pulse are transmitted with a phase difference of approximately  $\frac{\pi}{2} \; ({\rm mod}\,  \pi)$, such that their maximum possible power combining is minimized. 
Therefore, the best $L$ in this case is found such that it does not have any effect on the phases and~\eqref{eq:pulsePhdiff} is as close to zero as possible, i.e. 
\begin{eqnarray}
L &=& \left \lfloor \frac{\lambda}{2}\Ndata-  \frac{\Ne-1}{2}\right\rceil \nonumber\\ 
&=& \left \lfloor \frac{\lambda}{2}\Nsc-  \frac{\Ne-\lambda-1}{2} \label{eq:Lpi2}
\right\rceil.
\end{eqnarray} 
where $\lambda$ is an integer and $\left \lfloor \cdot \right\rceil$ returns the closest integer.  
With $\lambda=2$ and $\Ne$ even, then one can select $L= \Ndata-\frac{\Ne}{2}$ corresponding to symmetrical double-side SE considered in~\cite{Nokia21}. 
 
\emph{For QPSK and other QAMs}, choosing  
$L$ as in~\cite{Nokia21} 
is, however, sub-optimal, and most other $L$ values can provide slightly better PAPR. 
In these cases, the value of $L$ providing the lowest PAPR depends on the spectrum extension size $\Ne$ but also on the total bandwidth size via $\Ndata$ or $\Nsc$. Here, the lowest PAPR is obtained by creating a phase difference between neighboring pulses close to $\pi/4 \; ({\rm mod}\,  \pi/2)$. Namely, one should have
$
-\frac{2\pi}{\Ndata}  \left(L+\frac{(\Ne-1)}{2}\right)\approx \frac{\pi}{4}  \; \left({\rm mod}\,  \frac{\pi}{2}\right).
$ 
Therefore, the lowest PAPR should be achieved by selecting $L$ satisfying
\begin{eqnarray}
L&=& \left \lfloor \frac{(2\lambda+1)}{8} \Ndata-\frac{\Ne-1}{2}\right\rceil  \nonumber \\ 
&=& \left \lfloor  \frac{(2\lambda+1)}{8} \Nsc -\frac{(2\lambda+5)}{8} \Ne + \frac{1}{2} \right\rceil \label{eq:Lpi4}
\end{eqnarray}
where $\lambda$ is an integer.

	\begin{figure}[t]
\vspace{-0.0cm} 
\includegraphics[width=.5\textwidth]{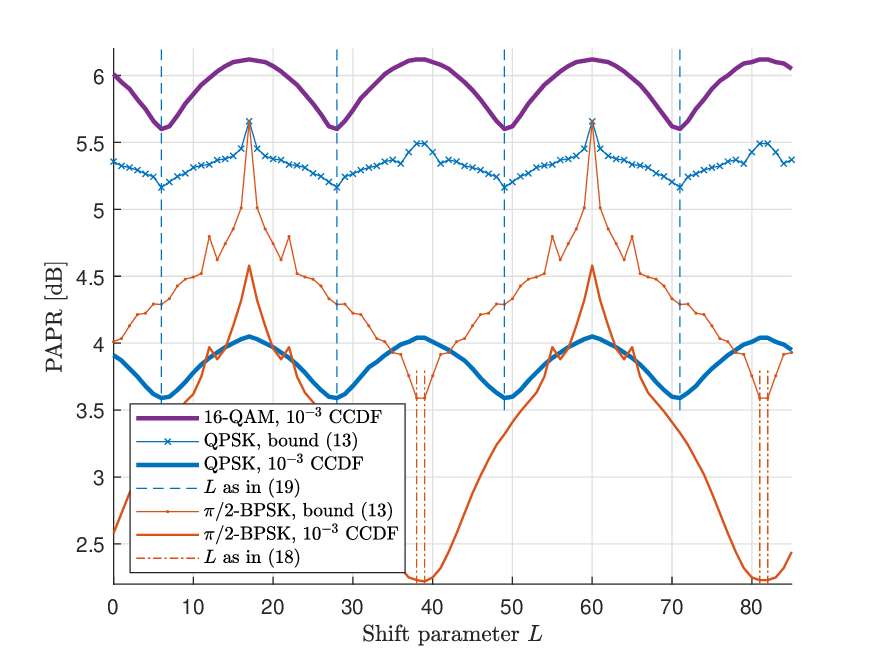}
\vspace{-0.7cm}
	\caption{Example of PAPR fluctuation as a function of different frequency-shift value $L$ for $\pi/2$-BPSK, QPSK and 16-QAM. 
	\label{fig:PAPRvsL}}
\vspace{-0.4cm}
\end{figure}

\begin{figure}[t]
\vspace{-0.0cm} 
\includegraphics[width=.5\textwidth]{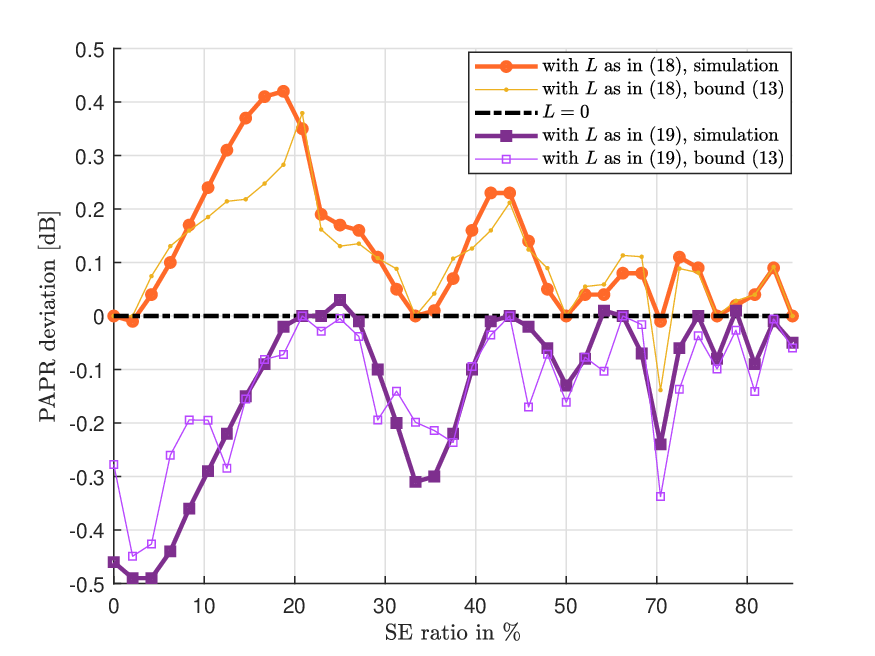}
\vspace{-0.7cm}
	\caption{PAPR deviation  of QPSK for different $L$ as a function of SE ratio $\Ne/\Nsc$. 
	\label{fig:PAPRgainFromL}
	}
\vspace{-0.3cm}
\end{figure}

Fig.~\ref{fig:PAPRvsL} shows the $10^{-3}$-CCDF PAPR  for $\pi/2$-BPSK, QPSK, and 16-QAM as a function of $L$ with $\Nsc =96$, $\Ne=10$. The FDSS is the deformed Hann window with  $\beta =-11 $ dB. The bounds~\eqref{eq:PAPRub} for  $\pi/2$-BPSK and QPSK are also shown.  The PAPR curves and their bounds have several equivalent minima as predicted by~\eqref{eq:Lpi2} for $\pi/2$-BPSK and by~\eqref{eq:Lpi4} for both  QPSK and 16-QAM. Note that for $\pi/2$-BPSK, the decimal part of~\eqref{eq:Lpi2} is $0.5$ so rounding up or down to the nearest integer is valid.  
One can observe that the choice of $L$ can have a large impact on the PAPR for  $\pi/2$-BPSK, but this impact is more moderate for higher-order QAM.
 
In Fig.~\ref{fig:PAPRgainFromL}, we compare the PAPR of QPSK with different values of $L$  as a function of the SE ratio $\Ne/\Nsc$, using the $10^{-3}$-CCDF PAPR with $\Nsc =96$ and  Kaiser FDSS window with $\kappa = 2 $ (maximum power attenuation of $7.15$ dB). 
The PAPR with $L$ as in~\eqref{eq:Lpi2} and $L$ as in~\eqref{eq:Lpi4}  are compared by  deviation from the one with $L=0$.  
 The figure also show a prediction of this PAPR deviation using the bound~\eqref{eq:PAPRub} with the corresponding $L$ values. As seen, $L$ as in~\eqref{eq:Lpi2}, which corresponds to the symmetric extension of~\cite{Nokia21}, when used with QPSK is always worse than $L=0$, which corresponds to the single-side extension in~\cite{SlimaneTVT07}. Similar results can be obtained for higher-order QAMs. The maximum PAPR gain over~\cite{Nokia21} by using a different $L$ is only a fraction of a dB, here up to 0.5 dB, and this gain  decreases as the SE size increases.

\subsection{Optimum SE Size $\NePAPR$  for Minimum PAPR}


\subsubsection{Observations on the Existence of an Optimal SE Size} 

\paragraph{Intuitive Explanation} Consider the pulse-shaping filters in~\eqref{eq:pm}, 
derived from the band-limited FDSS window $W[k]$  in~\eqref{eq:h}. Their amplitude essentially follows a sinc shape, with more or less attenuated side lobes. The kernel filter shape in \eqref{eq:h} is independent of $\Ne$,  
except for a scaling factor $\frac{1}{\sqrt{\Nsc-\Ne}}$, which  causes the peak energy of each pulse to increase as $\Ne$ increases.
In parallel, since there are $(\Nsc-\Ne)$ pulses per OFDM symbol, the pulses are spaced $\frac{\Nfft}{\Nsc-\Ne} $  samples apart in~\eqref{eq:pm}. This spacing  increases with $\Ne$, which in turn changes the degree of pulse overlap due the sinusoidal shape of the pulses. Both effects combined, depending of  $\Ne$, their overall combining may decrease or increase as a function of $\Ne$, leading to a non-monotonic PAPR variation.  
 
\begin{figure}[t]
\vspace{-0.0cm} 
\includegraphics[width=.5\textwidth]{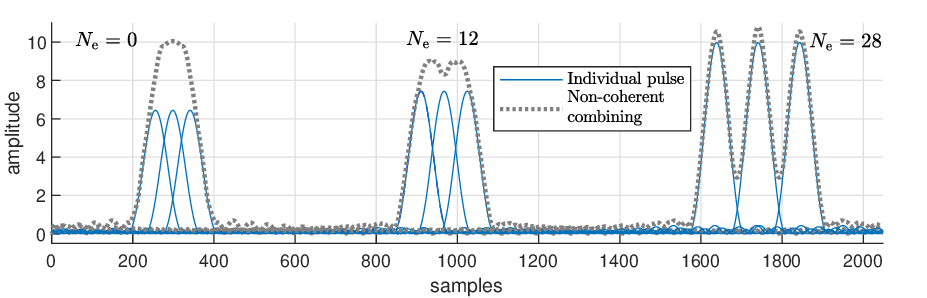}
\vspace{-0.7cm}
	\caption{Groups of 3 pulses and their non-coherent combining for different values of $\Ne$.   \label{fig:Pulses}}
\vspace{-0.5cm}
\end{figure}

In Fig.~\ref{fig:Pulses}, 	we illustrate this with a deformed Hann FDSS window ($\beta = -11$ dB and $\Nsc =48$). Three groups of three pulses are shown for different $\Ne$,  along with their respective non-coherent combining. 
Increasing the SE size to $\Ne=12$ raises the individual pulse energy but also spreads the pulses farther apart, reducing overlap and lowering their combined peak.   
However, further increasing to  $\Ne=28$ causes the larger individual pulse energy to dominate, raising the combined peak beyond the level without SE ($\Ne =0$). 

\begin{figure*}[t]
\centering
\vspace{-0.0cm} 
\subfigure[$\Nsc = 96$, No FDSS, QPSK \label{fig:PAPRvsNe_8PRB_L=0_NoFDSS_QPSK}]{\includegraphics[width=.32\textwidth]{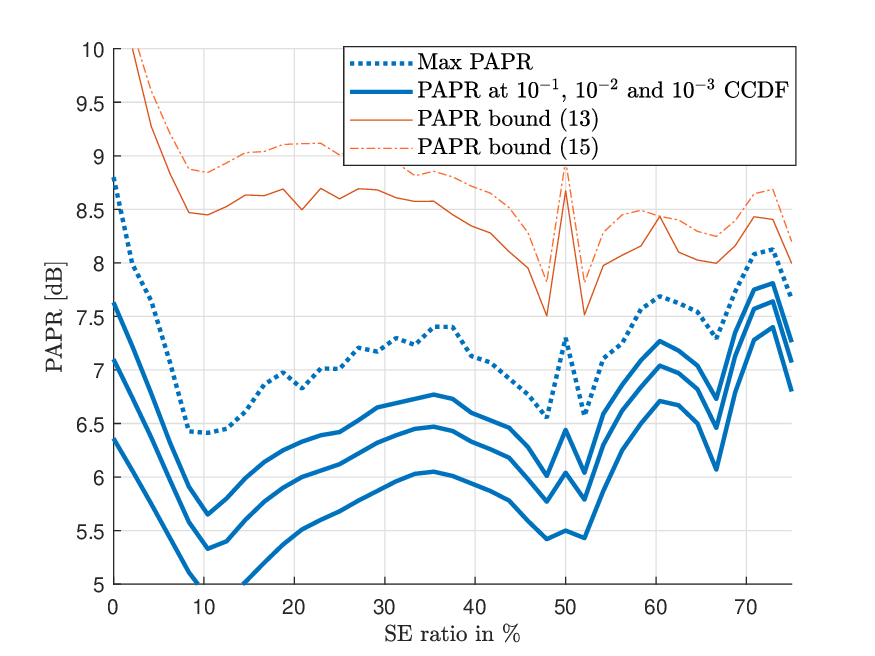}}
\subfigure[$\Nsc = 96$, Kaiser $\kappa =2$, QPSK \label{fig:PAPRvsNe_8PRB_L=0_Kaiser2_QPSK}]{\includegraphics[width=.32\textwidth]{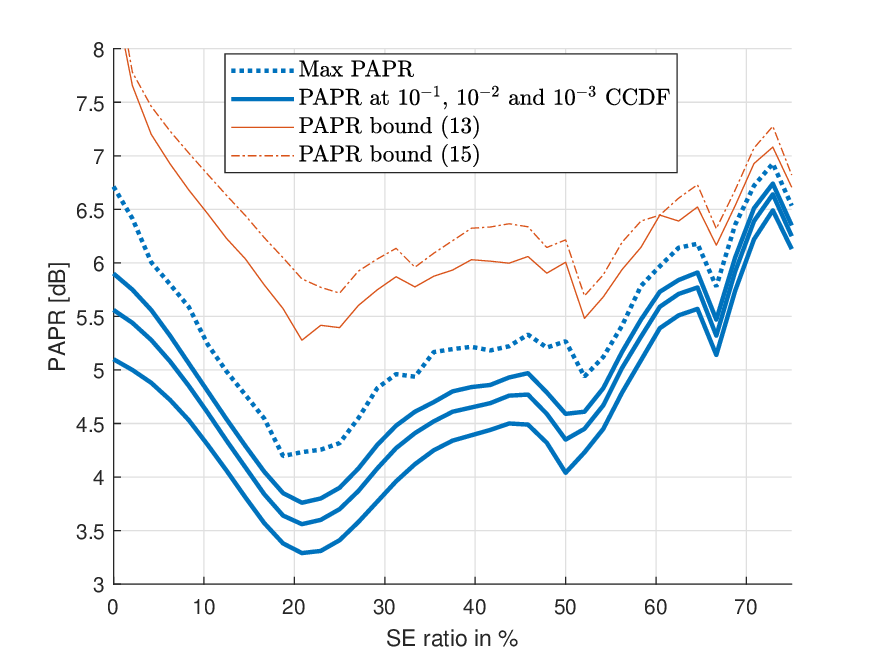}}
\subfigure[$\Nsc = 96$, Hann  $\beta =-11$ dB, QPSK\label{fig:PAPRvsNe_8PRB_L=0_Cos11_QPSK}]{\includegraphics[width=.32\textwidth]{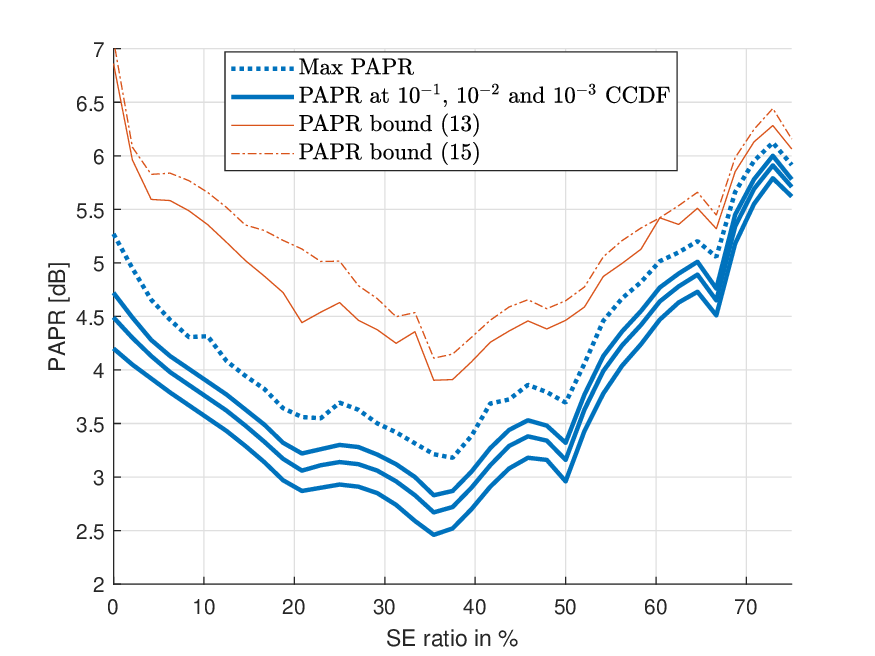}}
\subfigure[$\Nsc = 96$, Kaiser $\kappa =2$, $\pi/2$-BPSK \label{PAPRvsNe_8PRB_L=pi2_Kaiser2_pi2-BPSK}]{\includegraphics[width=.32\textwidth]{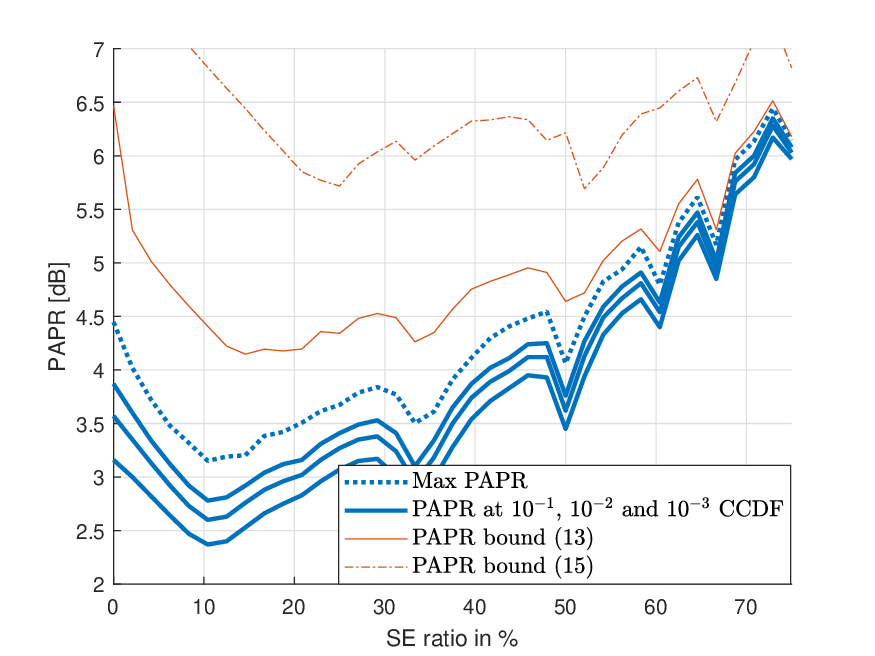}}
\subfigure[$\Nsc = 96$, Kaiser $\kappa =2$, 16-QAM \label{PAPRvsNe_8PRB_L=0_Kaiser2_16QAM.eps}]{\includegraphics[width=.32\textwidth]{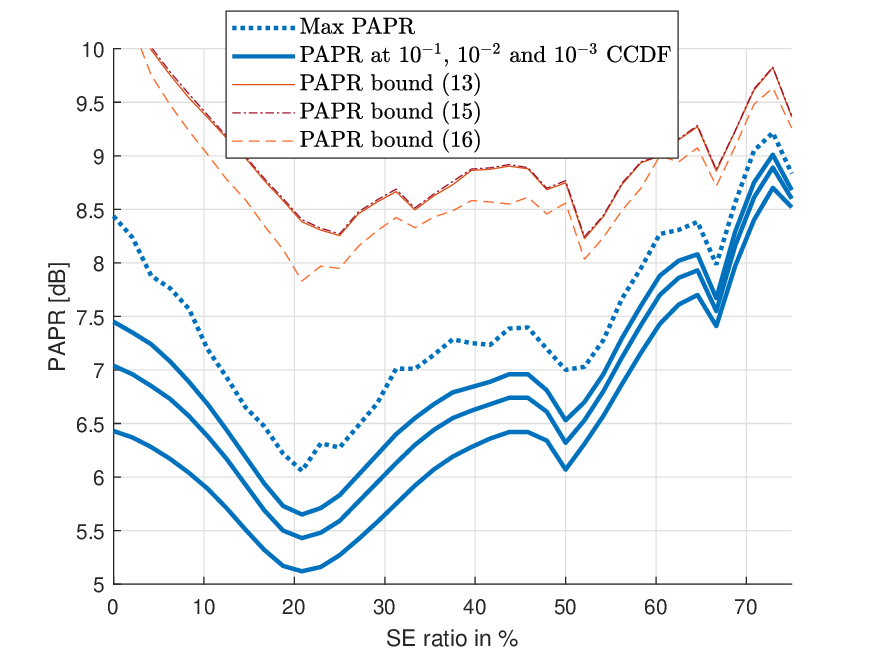}}
\subfigure[$\Nsc = 48$, Kaiser $\kappa =2$,  QPSK\label{PAPRvsNe_4PRB_L=0_Kaiser2_QPSK.eps}]{\includegraphics[width=.32\textwidth]{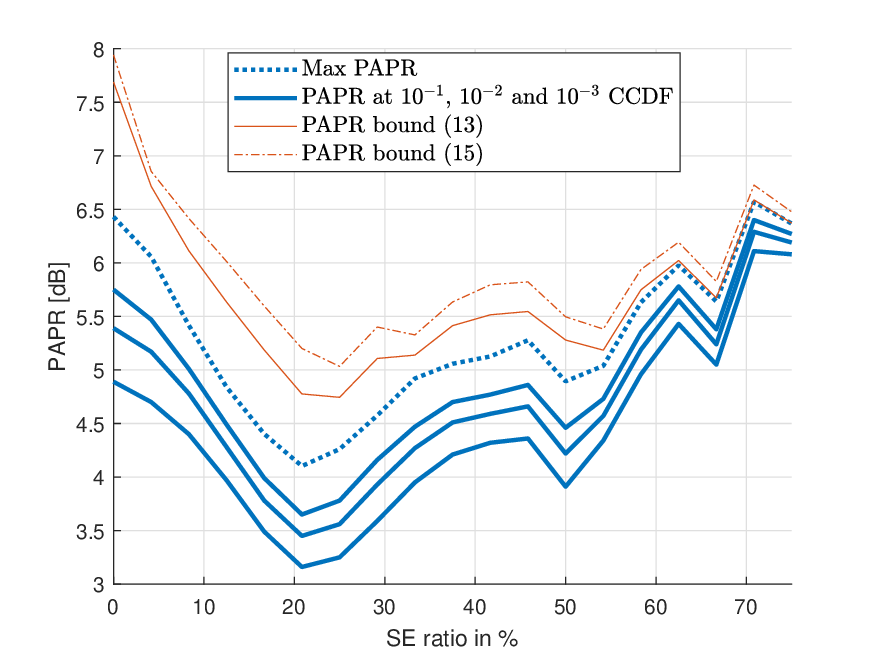}}
\vspace{-0.2cm}
	\caption{Examples of PAPR as a function of SE $\Ne$ showing the eixtence of a non-trivial SE size $\NePAPR$ for PAPR minimization. \label{fig:PAPRvsNe}}
\vspace{-0.2cm}
\end{figure*}

\paragraph{PAPR Simulations} 
The existence of this global optimum SE size that minimizes the PAPR can be systematically verified by simulations across many scenarios.  
In Fig.~\ref{fig:PAPRvsNe}, we plot several examples of PAPR  versus the SE ratio $\Ne/\Nsc$. Each plot shows the maximum  PAPR from $10^6$ Monte Carlo points, the $10^{-1}$, $10^{-2}$, and $10^{-3}$ CCDF, and the bounds from~\eqref{eq:PAPRub} and~\eqref{eq:PAPRub2}. 
All subplots is with QPSK, except Fig.~\ref{PAPRvsNe_8PRB_L=pi2_Kaiser2_pi2-BPSK} with $\pi/2$-BPSK and Fig.~\ref{PAPRvsNe_8PRB_L=0_Kaiser2_16QAM.eps} with 16QAM. All have $\Nsc = 96$ except Fig.~\ref{PAPRvsNe_4PRB_L=0_Kaiser2_QPSK.eps} with $\Nsc = 48$. 
The first row, i.e. Figs.~\ref{fig:PAPRvsNe_8PRB_L=0_NoFDSS_QPSK}--(c),  
uses different FDSS windows, while the second row, i.e. Figs.~\ref{PAPRvsNe_8PRB_L=pi2_Kaiser2_pi2-BPSK}--(f), varies the modulation or bandwidth, all with a Kaiser FDSS window ($\kappa=2$). Shift values match single-side extension ($L=0$) in the literature, except Fig.~\ref{PAPRvsNe_8PRB_L=pi2_Kaiser2_pi2-BPSK} where $L$  is chosen according to~\eqref{eq:Lpi2}.  

 Fig.~\ref{fig:PAPRvsNe} clearly shows that the PAPR does not decrease monotonically with $\Ne$.   
In all cases, the PAPR decreases up to an optimum SE size $\NePAPR$, beyond which it increases again.  
Selecting $\Ne> \NePAPR$ is detrimental not only for spectral efficiency but also for PAPR reduction. Remark that even without FDSS [Fig.~\ref{fig:PAPRvsNe_8PRB_L=0_NoFDSS_QPSK}], SE can provide some PAPR reduction and so $\NePAPR> 0$.   

From Figs.~\ref{fig:PAPRvsNe_8PRB_L=0_NoFDSS_QPSK}--(c) one can note that  $\NePAPR$ changes with the FDSS window, while the PAPR curves are almost identical between QPSK [Fig.~\ref{fig:PAPRvsNe_8PRB_L=0_Kaiser2_QPSK}] and 16-QAM [Fig.~\ref{PAPRvsNe_8PRB_L=0_Kaiser2_16QAM.eps}]. Similarly, the PAPR behavior seems consistent for QPSK  with different bandwidths [Figs.~\ref{fig:PAPRvsNe_8PRB_L=0_Kaiser2_QPSK} and~\ref{PAPRvsNe_4PRB_L=0_Kaiser2_QPSK.eps}].   


\subsubsection{Semi-Analytical Characterizations of $\NePAPR$ from Bounds}
Fig.~\ref{fig:PAPRvsNe} shows good agreement in the overall trend between the simulated PAPR  and the PAPR bounds~\eqref{eq:PAPRub} and~\eqref{eq:PAPRub2} as a function of $\Ne$. 
Accordingly, these bounds can be used to identify $\NePAPR$.  While an analytical determination of the bound minimizer  appears  intractable, its numerical evaluation is significantly less complex than direct PAPR simulations. This, in turn, 
facilitates the study of the optimum SE as a function of other system parameters.

Compared to~\eqref{eq:PAPRub2}, the bound~\eqref{eq:PAPRub} depends on the complete constellation, as well as $L$. The latter appears thus slightly more  precise on Fig.~\ref{fig:PAPRvsNe} for approximating $\NePAPR$. For Fig.~\ref{PAPRvsNe_8PRB_L=0_Kaiser2_16QAM.eps} with 16QAM, we also display the approximate bound~\eqref{eq:PAPRubQPSK} which appears to be a slightly better and simpler way to approximate  $\NePAPR$ than \eqref{eq:PAPRub} in this case. 
A defect of these bounds is however that the global optimum SE size obtained from simulations may correspond only to a local minimium in the PAPR bound, as shown in Fig.~\ref{fig:PAPRvsNe_8PRB_L=0_NoFDSS_QPSK} -- something   
we observed for smaller bandwidths and FDSS windows with small attenuation. A simple way to circumvent this issue is to restrict the search range for small FDSS attenuation, or inject a correction factor into the bound. Indeed, we observe that the bounds are further apart for $\Ne=0$ and become tighter as $\Ne \to \Nsc$. A practical method to compensate for this discrepancy is to approximate the PAPR as  
${\rm PAPR } \; {\rm [dB] } \approx  {\rm PAPR }^{\rm U}(\Ccal,\Ne,L)\; {\rm [dB] } - K \left(1-\frac{\Ne}{\Nsc}\right)$ 
where $K $ is the gap between the bound and simulation at $\Ne=0$, which can empirically be evaluated. 

\begin{figure*}[t]
\centering
\vspace{-0.2cm} 
\subfigure[Different QAM orders \label{fig:SEoptForPAPRdifferentQAM}]{\includegraphics[width=.45\textwidth]{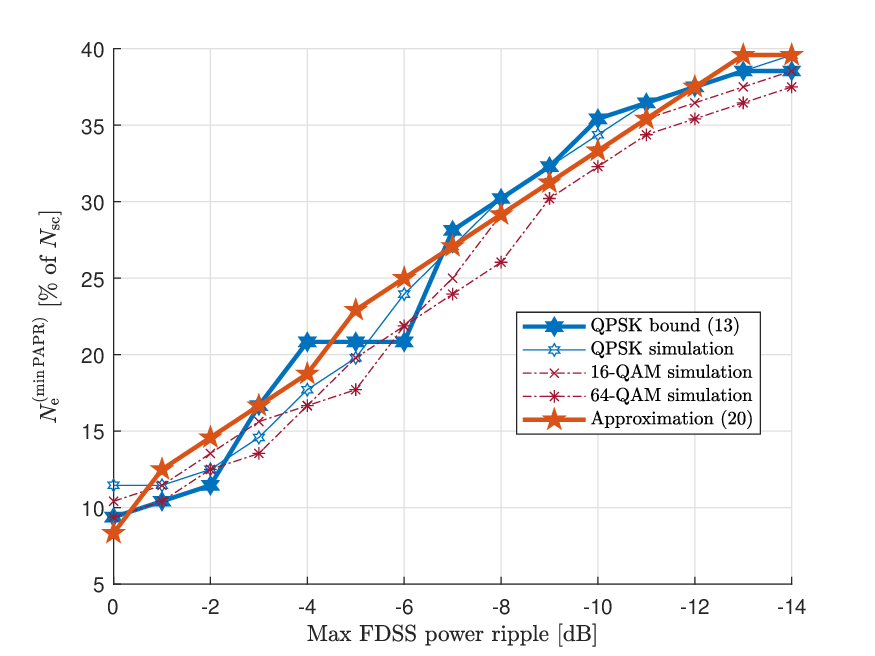}}
\subfigure[Different $L$ for QPSK and $\pi/2$-BPSK \label{fig:SEoptForPAPRdifferentL}]{\includegraphics[width=.45\textwidth]{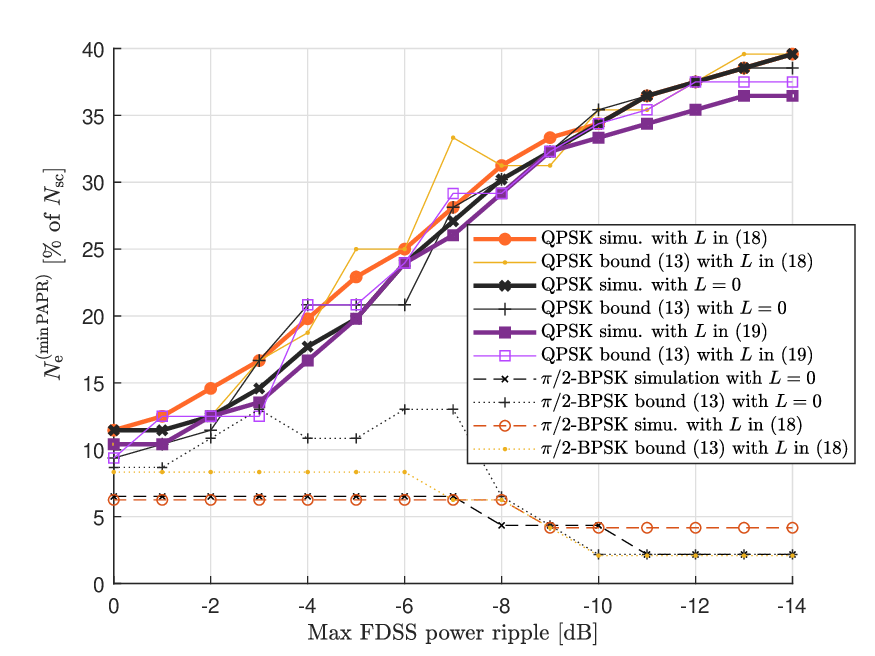}}
\vspace{-0.2cm}
	\caption{Optimum SE size $\NePAPR$  for different FDSS windows. \label{fig:SEoptForPAPR}}
\vspace{-0.0cm}
\end{figure*}

Since the search space and summations in ~\eqref{eq:PAPRub} and~\eqref{eq:PAPRub2} depend on the bandwidth size,  even numerically optimizing these bounds becomes computationally expensive when the bandwidth is large. 
We therefore consider a further simplification of the bound~\eqref{eq:PAPRub2}. 
As illustrated in Fig.~\ref{fig:Pulses}, 
once $\Ne$ increases and the pulses spread apart, 
the dominant contribution to the peak of the non-coherently combined pulses arises mainly from the overlap of two neighboring pulses, while the sidelobes of the remaining pulses are negligible, especially for larger FDSS attenuation. 
Accordingly, we reduce the summation in~\eqref{eq:PAPRub2} to two pulses as 
 $\max_n  \left\{ \sum_{i,j= 0}^{\Ndata-1} |p_i[n]||p_j[n]| \right\}  
\approx  \max_n  \left\{ \sum_{i,j= 0}^{1} |p_i[n]||p_j[n]| \right\}  
\!= \!\max_n  \left\{  (|p_0[n]| + |p_1[n]|)^2 \right\}
$. 
Since the absolute amplitude level is irrelevant for determining the minimizing location, the square exponent can  be omitted. With a  small reformulation of the kernel pulse, we thus obtain the approximation 
\begin{multline}
\NePAPR \approx \\
\argmin_{\Ne}  \frac{\max_n  \left\{  w[n]  + w\left[n-\frac{1}{\Nsc - \Ne}\right] \right\}}{\sqrt{\Nsc - \Ne}} \label{Eq:2PulsesApprox}
\end{multline}
where $w[n]= \sqrt{\Ndata} |p_0[n]| = \left| \sum_{k=0}^{\Nsc-1} W[k] e^{\jrm \frac{2 \pi k}{\Nfft}n}  \right|$ is the magnitude of the $\Nfft$-IDFT of the zero-padded FDSS window.

Solving~\eqref{Eq:2PulsesApprox} analytically appears still intractable, even in the absence of FDSS where the kernel reduces to the  closed-form Dirichlet kernel in~\eqref{eq:p0sinc}. Nevertheless, \eqref{Eq:2PulsesApprox} provides a simplified connection between the relevant parameters and the PAPR-minimizing SE size, while being computationally very efficient to evaluate.

\subsubsection{Comparison with Simulated $\NePAPR$}
\begin{figure}
\vspace{-0.0cm} 
\includegraphics[width=.5\textwidth]{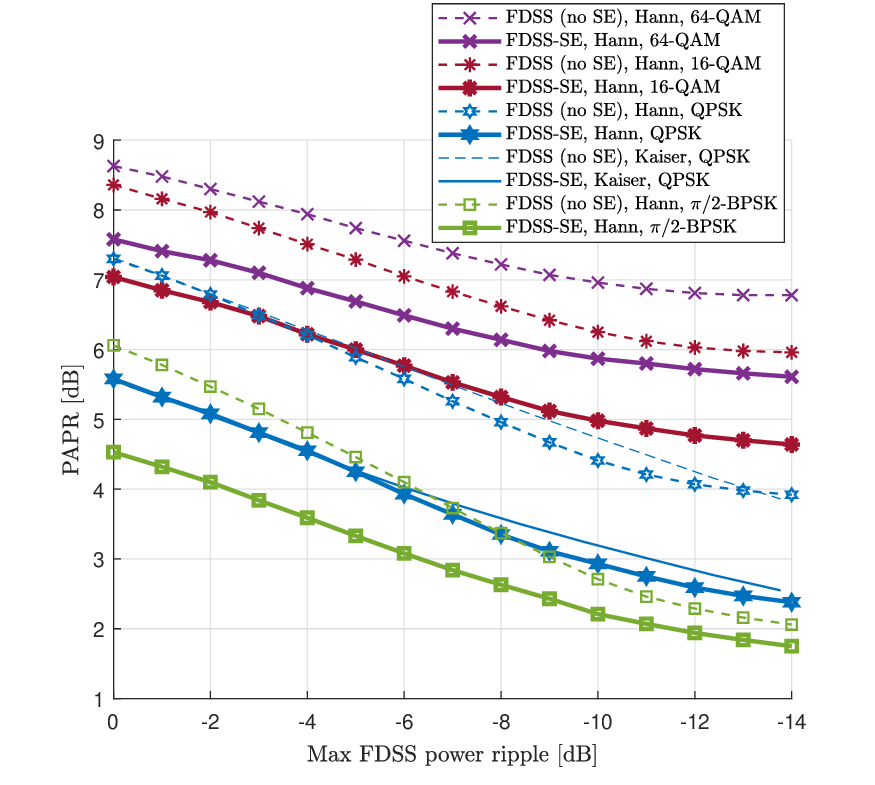}
\vspace{-0.6cm}
	\caption{Comparisons of the minimum PAPR achieved with and without SE over different FDSS attenuation.   \label{fig:OptPAPR}}
\vspace{-0.2cm}
\end{figure}

In Fig.~\ref{fig:SEoptForPAPRdifferentQAM}, we compare $\NePAPR$  obtained from simulations minimizing the $10^{-2}$ PAPR CCDF for  4-, 16-, and 64-QAM, with $\NePAPR$ estimated from the QPSK PAPR bound~\eqref{eq:PAPRub}. The range of $\NePAPR$ is $10$--$40\%$. As confirmed again, the optimum SE size $\NePAPR$ is largely independent of the QAM order which can be well approximated using the bound~\eqref{eq:PAPRub} or approximation~\eqref{Eq:2PulsesApprox}, with little impact on the resulting PAPR value\footnote{See the PAPR values for QPSK in Figs.~\ref{fig:OptPAPR} using the simulation-based estimation of $\NePAPR$, and the PAPR values in Fig.~\ref{fig:OptPAPR_vsPAPRNcapa_Hann} using the bound-based estimation of $\NePAPR$.}. Here, $L=0$ for all cases. 

 In Fig.~\ref{fig:SEoptForPAPRdifferentL}, we show that while the shift value of $L$ can impact the PAPR, the optimal SE size is nearly the same for different $L$ values. 
For QPSK, $\NePAPR$ with  $L$ as in~\eqref{eq:Lpi2},~\eqref{eq:Lpi4} and $L=0$ is obtained from simulations, together with the corresponding bound-based estimates. All cases are closely overlapping, and thus the bound~\eqref{eq:PAPRub} does not capture very well the small variation of $\NePAPR$ with different $L$ values. In Fig~\ref{fig:SEoptForPAPRdifferentL}, $\NePAPR$ is also shown for $\pi/2$-BPSK with $L$ as in~\eqref{eq:Lpi2}  and $L=0$, from simulation and corresponding bounds. In this case, the range of $\NePAPR$ is much smaller and almost constant around $5\%$,  indicating that only a small amount of SE may be beneficial. Moreover, contrary to QPSK and other QAMs, as FDSS shaping increases, $\NePAPR$  decreases. 

\begin{figure}[t]
\vspace{-1cm} 
\includegraphics[width=.5\textwidth]{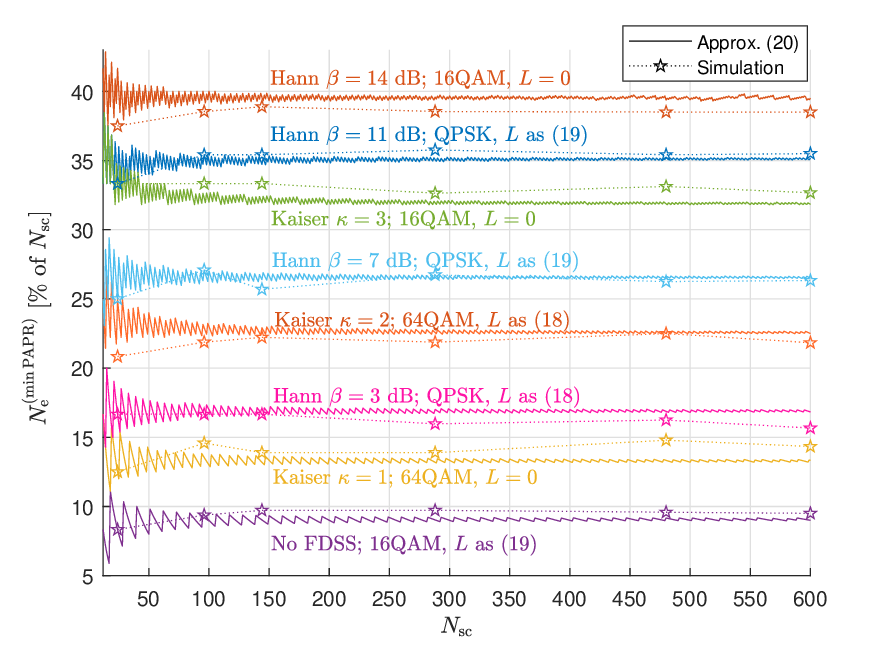}
\vspace{-0.8cm}
	\caption{Optimum SE size as function of bandwidth size.   \label{fig:NeOptPAPR_vs_2PulsesApprox}}
\vspace{-0.2cm}
\end{figure}

In Fig.~\ref{fig:NeOptPAPR_vs_2PulsesApprox}, we compare the simulated   $\NePAPR$ for different bandwidth sizes $\Nsc=24,\,96,\,144,\,288,\,480$, and $600$  with  the approximation~\eqref{Eq:2PulsesApprox}. Various configurations of QAM order, frequency shift $L$, and FDSS window are considered. Here, $\Nfft = 8192$, and the simulations target the $10^{-2}$-CCDF PAPR. 
Both simulations and the approximation in~\eqref{Eq:2PulsesApprox} show that, as the bandwidth increases, the value of $\NePAPR$ approaches a constant fraction of the bandwidth size.  The sawtooth-shaped behavior in~\eqref{Eq:2PulsesApprox} stems from the integer constraint on $\Ne$.  
As $\Nsc$ increases, the resolution of $\Ne/\Nsc$ improves, and the optimum SE ratio converges to  a precise spacing between pulses. 

\subsection{Simulation Evaluations of Maximum PAPR gain of FDSS-SE over FDSS} 
Finally, Fig.~\ref{fig:OptPAPR} shows the $10^{-3}$ PAPR CCDF  as a function of FDSS shaping, both with and without SE. To reach the lowest PAPR values, the shift parameter $L$ is according to~\eqref{eq:Lpi2} for $\pi/2$-BPSK and~\eqref{eq:Lpi4} for others; and for FDSS-SE, the corresponding optimum SE sizes  $\NePAPR$  estimated by simulation are used.
For QPSK, both filter families -- Hann and Kaiser windows -- are considered, while for others only results with Hann window are shown for clarity. 

For QPSK, SE with Hann window provides an almost constant PAPR gain over FDSS without SE from $1.74$ dB to $1.54$ dB. With Kaiser window, the gain of SE is the same with no FDSS but decreases slightly more to $1.31$ dB. For both window types, the lowest PAPR achieved with FDSS-SE is approximately the same  for a given maximum  window attenuation ripple. 
For 16-QAM and 64-QAM, the PAPR gain with Hann window is also constant but smaller, at about $1.3$ and $1.1$ dB, respectively.   
For $\pi/2$-BSPK, the behavior is different: the gain of SE over FDSS gradually  decreases with larger shaping, from $1.5$dB to $0.3$dB. 

This is consistent with the consensus during 3GPP 5G studies that SE on top of FDSS provides little benefit for $\pi/2$-BSPK~\cite{R1-2300090}, especially since in 5G the FDSS windows have been allowed to have such levels of attenuation ripple. 
Overall FDSS-SE provides the greatest PAPR reduction potential for QPSK, reaching values as low as 2-3 dB --  similar to those obtained with $\pi/2$-BSPK using FDSS alone.

\section{Spectrum Extension Maximizing the Transmission Rate}

The PAPR reduction from FDSS comes at the cost 
 of self-induced inter-carrier interference (ICI) as the time-multiplex pulses are not orthogonal anymore, which reduces the transmission rate.   
The PAPR reduction obtained by SE comes at the cost of a smaller effective bandwidth, which can also degrade the transmission rate -- although, as we will show, not always. In this section, we discuss the achievable rate of FDSS-SE as a function of the SE size.

\subsection{DFT-s-OFDM Reception}
The transmitted signal $s[n]$ is assumed to be received over a $P$-path channel $\{ h_p\}_{p=0}^{P-1}$,  with $P< \Ncp$, such that the received OFDM symbol for $0\leq n \leq (\Nfft-1)$ is  
\begin{equation} \label{eq:y[n]}
y[n]= \sqrt{\snr}\sum_{l=0}^{P-1}{ h_p s[n-p]} + z[n].
\end{equation} 
where $\snr$ denotes the signal-to-noise ratio (SNR),  $\{z[n]\}$ is an additive white Gaussian noise (AWGN) with $\expect{|z[n]|^2 }=1$, and the channel is normalized so that  
$\sum_{l=0}^{P-1}\expect{|h_p|^2} =1 $.   

After FFT demodulation, the frequency-domain symbol on subcarrier $0\leq k\leq \Nsc-1$  is 
\begin{eqnarray} \label{eq:Y[k]}
 Y[k]&=& \frac{1}{\sqrt{\Nfft}} \sum_{n=0}^{\Nfft-1} y[n] e^{-\jrm 2 \pi \frac{kn}{\Nfft}}  \nonumber\\
  &=& H[k]  X^{\rm se}[k] +Z[k]	
\end{eqnarray}
with per-subcarrier channel coefficient 
\begin{equation}
H[k]=\sqrt{\snr} W[k]\bar{H}[k]
\end{equation} 
where $\bar{H}[k]=\sum_{p=0}^{P-1} h_p e^{-\jrm 2\pi\frac{k p}{\Nfft}} $ is the  DFT of the channel impulse response  
and $Z[k]= \frac{1}{\sqrt{\Nfft}} \sum_{n=0}^{\Nfft-1} z[n] e^{-\jrm 2 \pi \frac{kn}{\Nfft}}$ is  the frequency-domain noise with unit variance.

\subsubsection{Equalization} 
We adopt the mixed MRC-MMSE combining from~\cite{Nokia21} which was the primary approach\footnote{An alternative could be applying combining before equalization as $Y[k]+Y[k+\Ndata ]  =(H[k]+H[k+\Ndata ])X^{\rm se}[k] +Z[k]+Z[k+\Ndata]$
but this is worse as  $|H[k]+H[k+\Ndata ]|^2\leq |H[k]|^2+|H[k+\Ndata ]|^2$.
} considered in the 3GPP Rel-18 study on uplink coverage enhancement, generalized here to any shift value $L$. 
With perfect channel knowledge,   
the symbols are first equalized via MRC: 
\begin{equation} \label{eq:R[k]}
R[k] = H[k]^* Y[k]   \quad \text{ for }  0\leq  k < \Nsc 
\end{equation}
and then repeated symbols are combined as 
\begin{equation} \label{eq:Rt[k]}
\tilde{R}[k]= \begin{cases}  
R[k]+R[k+\Ndata]&    0\leq  k< \Ne  \\   
R[k] & \Ne  \leq  k< \Ndata                  
\end{cases}. 
\end{equation}
The combining in~\eqref{eq:Rt[k]} is equivalent to~\cite[Eq. (8)]{Nokia21}, but here it is simplified based on the symbol symmetry shown  in Fig.~\ref{fig:SE_illustration}, avoiding  the need to define  five sub-bands as in~\cite{Nokia21}.

By substitution, 
\begin{equation}
R[k]= |H[k]|^2 X^{\rm se}[k] +H[k]^* Z[k] 
\end{equation}
and using the periodicity $X^{\rm se}[k+\Ndata ]=X^{\rm se}[k]$, we obtain
\begin{eqnarray}
\tilde{R}[k]&=&G[k] X^{\rm se}[k]+\tilde{Z}[k] \quad \quad \quad \quad \text{ for }  0\leq  k < \Ndata \nonumber \\
&=&G[k] X[(k+L) \;{ \rm mod}\, \Ndata]+\tilde{Z}[k]  \label{eq:Rt[k]2}
\end{eqnarray}
where  
\begin{equation} \label{eq:G[k]}
G[k]= \begin{cases}  
|H[k]|^2+|H[k+\Ndata]|^2 &   0\leq  k< \Ne  \\   
|H[k]|^2 &               \Ne  \leq  k< \Ndata                  
\end{cases}.
\end{equation}
and   the equalized noise is  
\begin{multline} \label{eq:Ztilde[k]}
\tilde{Z}[k]= \\ 
\begin{cases}  
H[k]^*Z[k]+H[k+\Ndata]^*Z[k+\Ndata]\;   0\leq  k< \Ne  \\ 
H[k]^*Z[k]  \quad \quad\quad\quad \quad \quad\quad\quad       \Ne  \leq  k< \Ndata 
\end{cases}. 
\end{multline}

 Accordingly, the MMSE equalization of the combined symbols is  
 \begin{equation} \label{eq:Req[k]}
\tilde{R}_{\rm eq}[k]= \frac{\tilde{R}[k]}{G[k]+1}. 
\end{equation}

\subsubsection{DFT Despreading}
The frequency shift is then reverted as\footnote{Since this cyclic shift is equivalent to a constellation phase rotation, an alternative is to first apply the IDFT, followed by  linear phase compensation on the demodulated symbols.}
 \begin{equation} \label{eq:Reqp[k]}
\tilde{R}_{\rm eq}'[k]= \tilde{R}_{\rm eq}[k-L \;({\rm mod}\; \Ndata)]
\end{equation}
followed by the IDFT
 \begin{equation} \label{eq:r[m]}
r[m]= \frac{1}{\sqrt{\Ndata}} \sum_{k=0}^{\Ndata-1} \tilde{R}_{\rm eq}'[k] e^{\jrm \frac{2\pi}{\Ndata}km}.
\end{equation}

\subsection{Achievable Rate with FDSS-SE}
From Appendix~\ref{Proof:Prop1}, FDSS-SE yields the following achievable rate, that we will refer with a slight abuse of terminology as the FDSS-SE capacity.  
\begin{Lem}\label{Prop1}
The effective SINR of the demodulated symbol in \eqref{eq:r[m]} is independent of $m$ and given by 
 \begin{equation} \label{eq:SNReff}
\SINR^{\rm eff} = \frac{g_{0}}{1-g_{0}}
\end{equation}
where $g_{0}= \frac{1}{\Ndata} \sum_{k=0}^{\Ndata-1} \frac{G[k]}{G[k]+1}$ with combined channel gains $G[k]$ in~\eqref{eq:G[k]}, and the achievable rate is  
 \begin{equation} \label{eq:Capa}
C = \frac{\Ndata}{\Nsc} \expect{\log_2 \left( \frac{1}{1-g_{0}} \right)}   \quad {\rm [bpcu]}.
\end{equation}  
\end{Lem}

\begin{figure}[t]
\vspace{-0.2cm} 
\includegraphics[width=.45\textwidth]{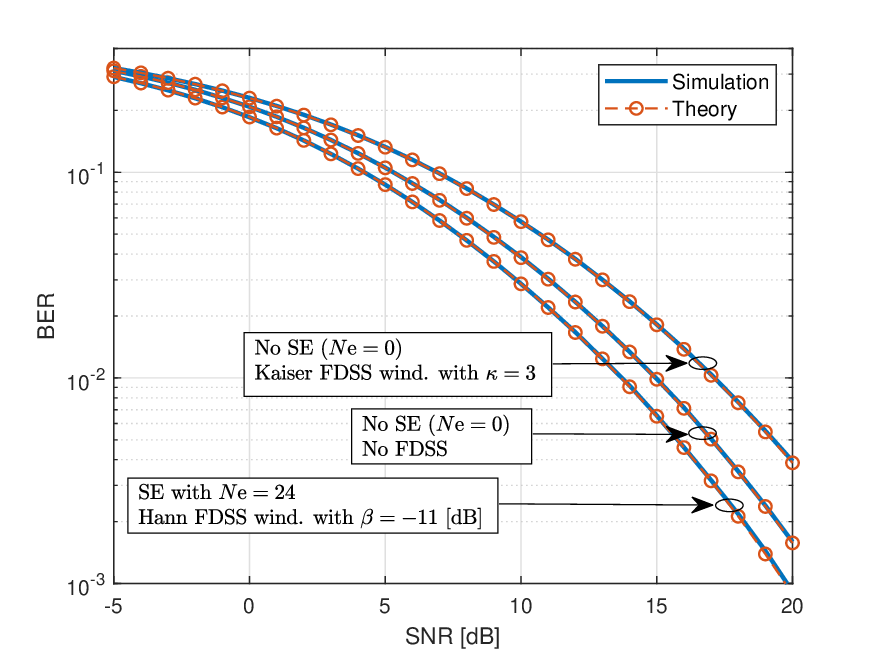}
\vspace{-0.2cm}
	\caption{Simulated BER of QPSK  versus theoterical BER using $\SINR^{\rm eff} $  in Lem.~\ref{Prop1}. \label{fig:BERThvsSim}}
\vspace{-0.3cm}
\end{figure}

With the intention of verifying Lem.~\ref{Prop1}, Fig.~\ref{fig:BERThvsSim} compares simulated bit error rates (BER)
of Gray-mapped QPSK with theoretical BER of QPSK  under $\SINR^{\rm eff} $ from~\eqref{eq:SNReff}, i.e., $Q\left(\sqrt{\SINR^{\rm eff}} \right)$ averaged over the different channel realizations. Channel assumptions are  $\Nsc = 96$, 3GPP TDL-C channel with $300$ ns delay spread and Rayleigh block fading. The near-perfect match confirms Lem.~\ref{Prop1}, indicating that the ICI behaves nearly Gaussian.

\begin{Rem}\label{RemBasicRx}
For  a basic receiver  without SE symbol combining as discussed in~\cite{Nokia21}, Lem.~\ref{Prop1} still applies but with  
$ G[k]= |H[\Ne/2+k]|^2 $ for $ 0\leq  k<  \Ndata$ instead of~\eqref{eq:G[k]}.
\end{Rem}

\begin{figure}[t]
\vspace{-0.2cm} 
\subfigure[RRC FDSS]{\includegraphics[width=.5\textwidth]{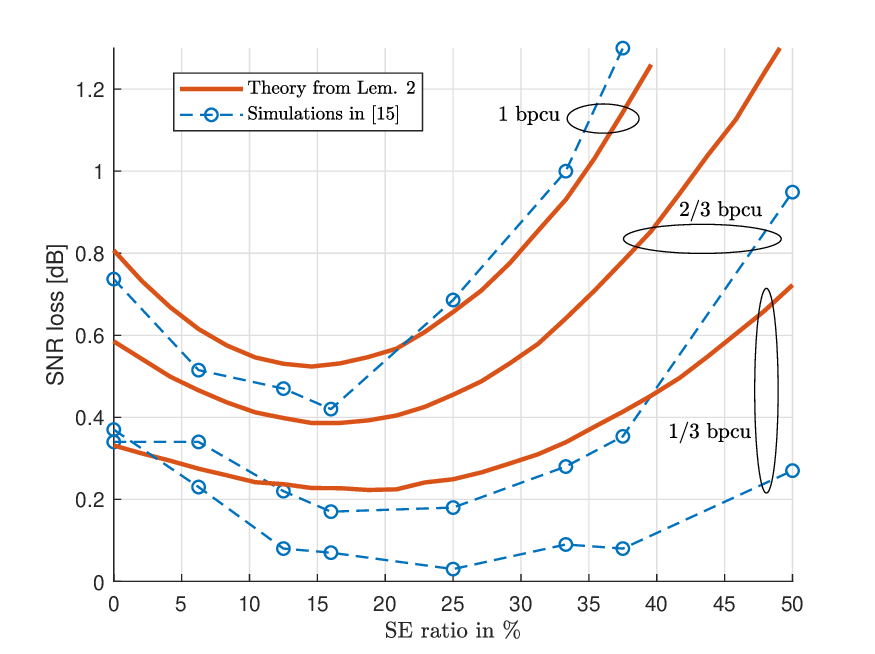}}
\subfigure[Hann FDSS]{\includegraphics[width=.5\textwidth]{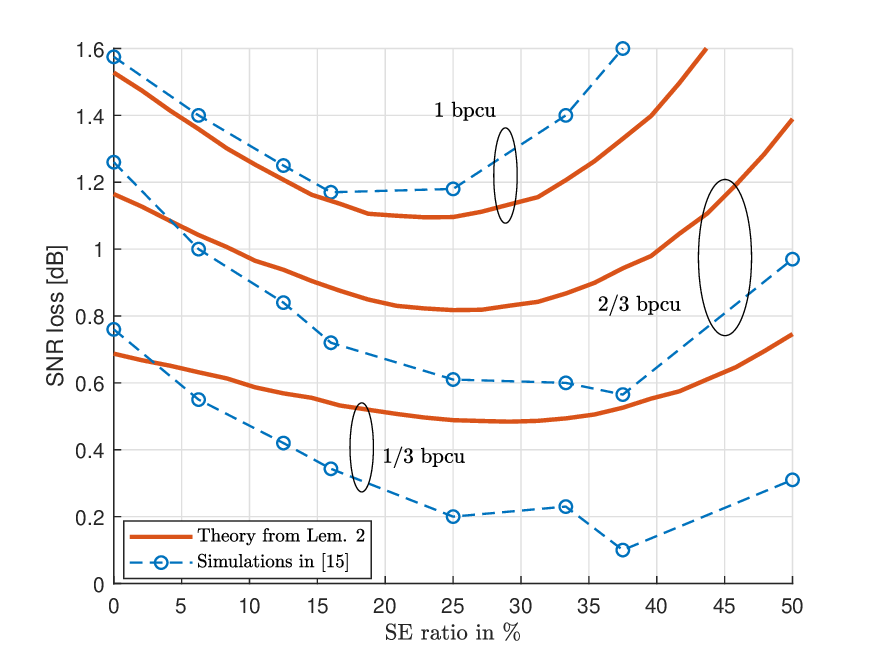}}
\vspace{-0.2cm}
	\caption{SNR loss of FDSS-SE wrt plain DFT-s-OFDM, at different spectral efficiency in [bpcu],  estimated either from the capacity formula  in Lem.~\ref{Prop1} or from the BLER simulation in~\cite{R1-2208412},  as a function of the SE ratio $\frac{\Ne}{\Nsc}$. \label{fig:SNRloss8RBs}}
\vspace{-0.2cm}
\end{figure}

\subsection{Optimum SE Size $\NeCapa$ for Maximum Rate} 
\subsubsection{Observations on the Existence of an Optimal SE} 
\paragraph{Intuitive Explanation} 
The FDSS-SE capacity~\eqref{eq:Capa} has a pre-logarithm scaling factor that decreases with SE size. 
However,  the term $g_0$ inside the logarithm may either decrease (due to a smaller summation  range) or increase (due to larger $G[k]$ values)  as a function of $\Ne$. Thus, as in the PAPR minimization case, 
 a non-obvious optimum  SE size $\NeCapa$ exists for capacity maximizing; and a too small SE size with  $\Ne <\NeCapa$ may not achieve the maximum rate.

\paragraph{BLER Simulations} A close examination of 3GPP BLER results confirms that FDSS-SE performance is non-monotonic with SE size. For example, rate improvements with SE can be observed in~\cite[Tabs. 2-7]{R1-2208412},  and the achievable-rate formulation in~\eqref{eq:Capa}  can help to predict such performance behavior in realistic setups. Fig.~\ref{fig:SNRloss8RBs} shows the SNR loss of FDSS-SE relative to plain DFT-s-OFDM versus the SE ratio for $\Nsc=96$ with 15 kHz subcarrier spacing, and a TDL-C(300 ns) channel. Two FDSS window types are considered: a 3-tap filter equivalent (up to half a subcarrier shift) to a deformed Hann window with $\beta=-11$ dB, and the TRRC window from~\cite{Nokia21} with parameters $(\rho,\beta)= (0.5,-0.65)$. The theoretical SNR loss is obtained by numerically finding the required SNR -- with and without FDSS-SE -- in~\eqref{eq:Capa} to achieve target spectral efficiencies of  $1/3$, $2/3$, and $1$ bpcu. This is compared with 3GPP evaluations of the SNR loss reported in~\cite[Tabs. 2-7]{R1-2208412} for reaching $10^{-1}$ BLER with QPSK and code rates $1/6$, $1/3$, and $1/2$. We note that these BLER simulations are not from the present authors, and follows the practical 3GPP assumptions in~\cite[Tab. 1]{R1-2208412}.  
As shown in  Fig.~\ref{fig:SNRloss8RBs}, the SNR loss  evaluated from 3GPP BLER evaluations in~\cite{R1-2208412} and the theoretical SNR loss from~\eqref{eq:Capa} form bell-shaped curves versus the SE ratio with consistent trends, particularly for half-rate coding (1 bpcu); such that, the theoretical optimal SE size $\NeCapa$ matches well the optimal SE size evaluated through BLER simulations. 

\subsubsection{Semi-Analytical Characterization of $\NeCapa$ from AWGN Capacity}
Finding the optimum SE size maximizing the capacity~\eqref{eq:Capa} is considerably simpler than optimizing BLER through full link-level simulations. Nevertheless, its evaluation still requires Monte-Carlo averaging to account for the effects of fading channels.  

In general, the optimal SE size for rate maximization depends primarily on the FDSS shaping, since the fluctuation in the combined channel gains $G[k]$ mainly depend on the FDSS. To isolate and highlight this dependency, we consider the no-fading AWGN channel case, for which the optimal SE size can be formulated as 
\begin{multline}
\NeAWGN = \argmax_{\Ne}   \left(1-\frac{\Ne}{\Nsc}\right) \log_2 \left( \frac{1}{1-g(\Ne)} \right) \label{eq:NeAWGN}
\end{multline}
where $g(\Ne)= \frac{1}{\Nsc-\Ne}\!\!\!\!\!\! \displaystyle \sum_{k=0}^{\Nsc-\Ne-1} \!\!\!\!\!\! \frac{1}{1+(\snr W^c_{\Ne}[k])^{-1}}$ 
with $W^c_{\Ne}[k]= |W[k]|^2+|W[k+\Nsc-\Ne]|^2$ for $ 0\leq  k< \Ne  $ and $W^c_{\Ne}[k]= |W[k]|^2 $ otherwise. 

The expression in~\eqref{eq:NeAWGN} can be efficiently evaluated numerically and reveals several expected behaviors of  $\NeCapa$. 
As  $\snr \to \infty$, $g(\Ne) \approx (1-f(\Ne)/\snr) $ for some bounded function $f(\Ne)$, such that the variation of $g(\Ne)$ with respect to $\Ne$ diminishes. Consequently, the AWGN capacity behaves as $C^{\rm AWGN} \approx \left(1-\frac{\Ne}{\Nsc}\right) \log_2 (\snr/f(\Ne)) $ implying that $\NeAWGN \to 0$ as  $\snr \to \infty$ so as to maximize  the pre-log multiplexing factor. 
In the special case of no FDSS, we obtain $g(\Ne) = \frac{1}{1+\snr^{-1}}\left(\Nsc - \frac{2\Ne}{2+\snr^{-1}} \right)$ which is a linearly decreasing function of $\Ne$, leading again to 
 $\NeAWGN = 0$. 
 Therefore, it can be generally expected that $\NeCapa$ decreases and tends to zero as SNR increases or as FDSS shaping diminishes.  Overall, $\NeCapa$ depends primarily on SNR and FDSS shaping.

\begin{figure}[t]
\centering
\vspace{-0.0cm} 
\subfigure[$\beta =11$dB, $\snr=5$dB, 30ns \label{fig:CapavsNe_Cos11_5DBsnr_30ns}]{\includegraphics[width=.24\textwidth]{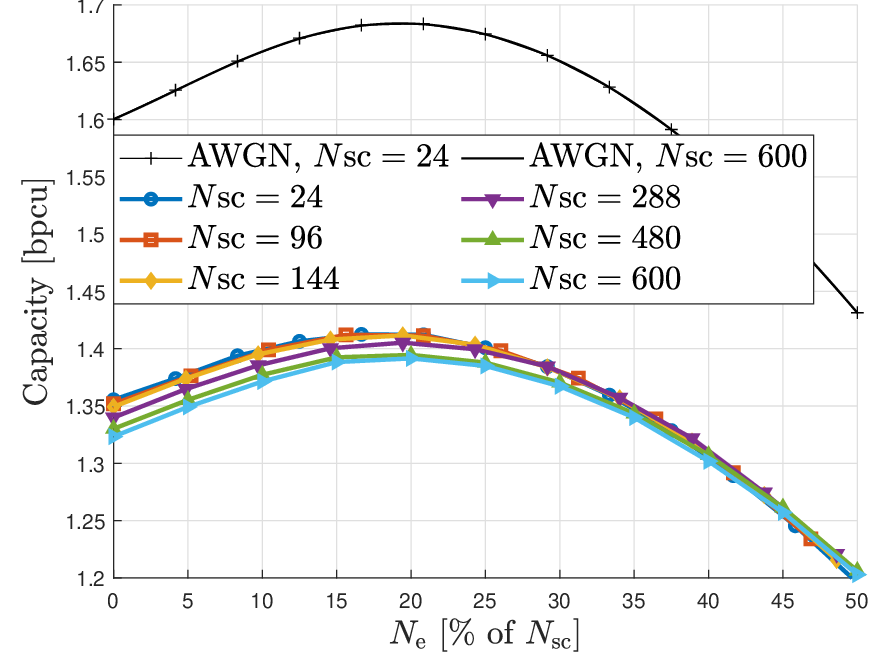}}
\subfigure[$\beta =11$dB, $\snr=5$dB, 300ns \label{fig:CapavsNe_Cos11_5DBsnr}]{\includegraphics[width=.24\textwidth]{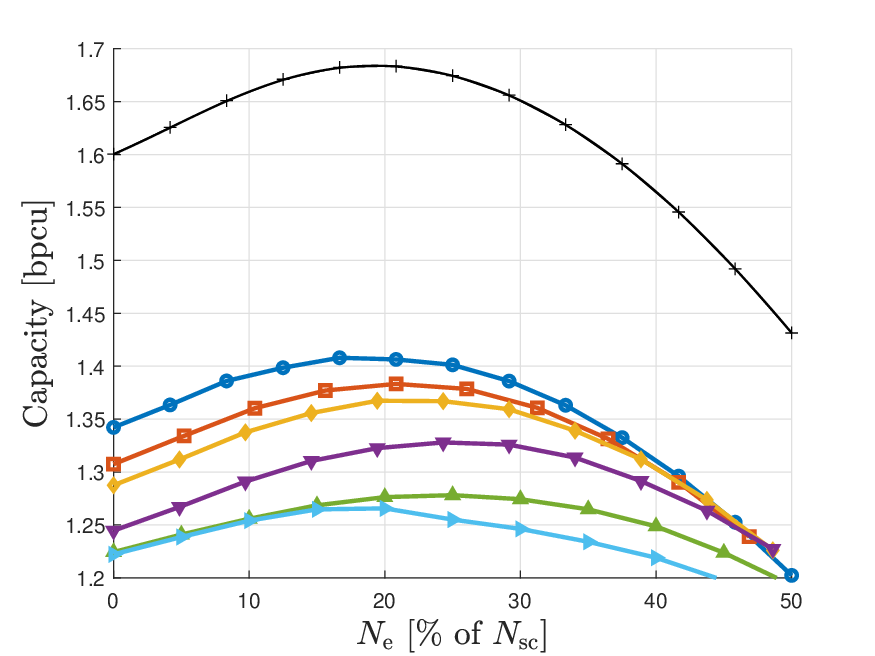}}
\subfigure[$\kappa =2$, $\snr=0$dB, 300ns\label{fig:CapavsNe_Kaiser2_0DBsnr}]{\includegraphics[width=.24\textwidth]{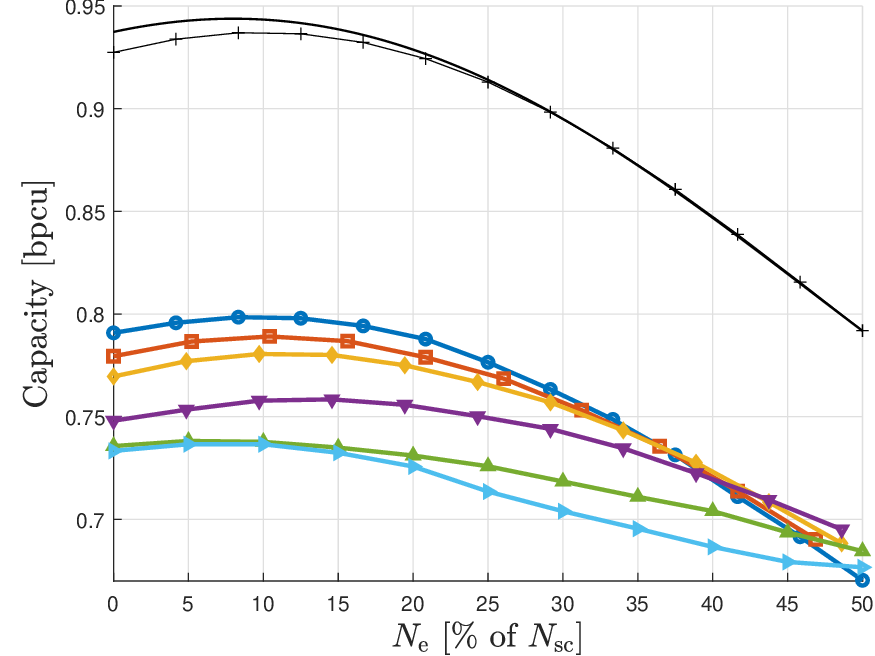}}
\subfigure[$\beta =11$dB, $\snr=0$dB,  300ns \label{fig:CapavsNe_Cos11_0DBsnr}]{\includegraphics[width=.24\textwidth]{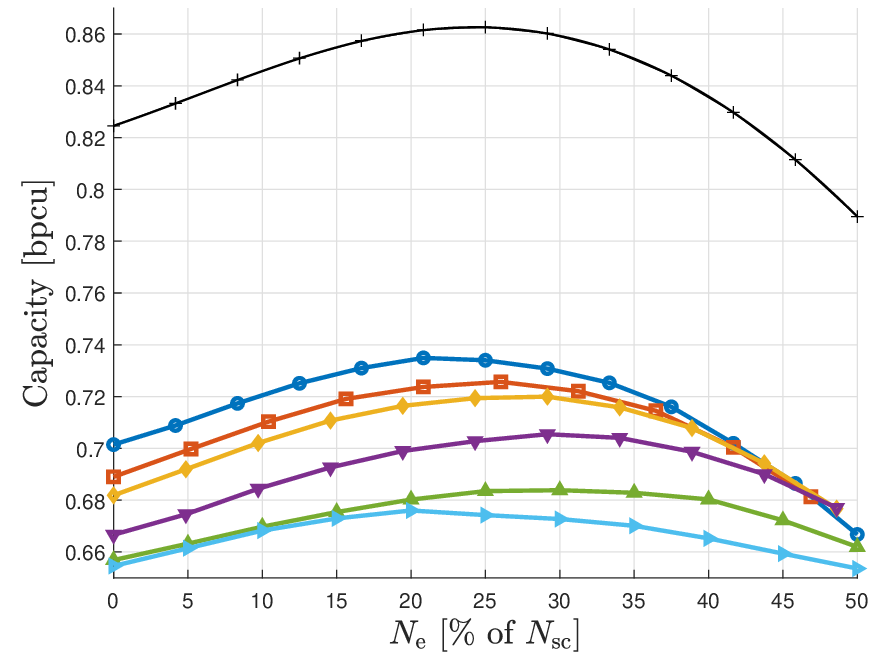}}
\vspace{-0.2cm}
	\caption{Capacities as a function of SE ratio $\Ne/\Nsc$ for different bandwidths $\Nsc$,  Hann ($\beta$) or Kaiser ($\kappa$) FDSS window, $\snr=0$ or 5 dB, and TDL-C channel with 30ns or 300ns delay spread.   
	\label{fig:CapavsNe}}
\vspace{-0.2cm}
\end{figure}

\begin{figure}[t]
\vspace{-0.0cm} 
\includegraphics[width=.5\textwidth]{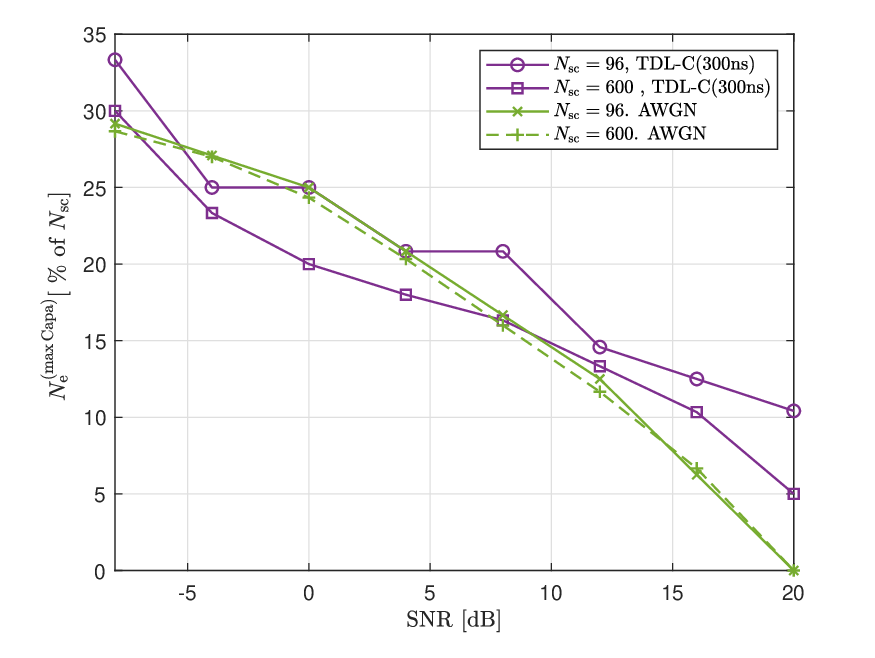}
\vspace{-0.8cm}
	\caption{Optimal SE size $\NeCapa$ in \% of $\Nsc$ for two channel models and two bandwidth size, as function of SNR, given a deformed Hann FDSS window with $\beta=-11$ dB.  \label{fig:NeCapOpt_vs_SNR_Cos11}}
\vspace{-0.0cm}
\end{figure}

\subsubsection{Comparison with Simulated $\NeCapa$}
Fig.~\ref{fig:CapavsNe} shows the achievable rates as a function of SE ratio $\Ne/\Nsc$ for different bandwidth sizes $\Nsc = 24, \, 144, \, 288, \, 480$ and $600$. 
Four configurations are considered. Three cases (Figs.~\ref{fig:CapavsNe_Cos11_5DBsnr_30ns}, \ref{fig:CapavsNe_Cos11_5DBsnr}  and~\ref{fig:CapavsNe_Cos11_0DBsnr}) use a Hann FDSS window ($\beta=-11$ dB), with $\snr =0$ or $5$ dB, and a TDL-C channel with either 30ns or 300ns delay spread. 
The remaining case considers a Kaiser window ($\kappa=2$) with $\snr=0$ dB and TDL-C with 300ns delay spread. AWGN capacities for $\Nsc = 24$ and $600$ are shown as references. The AWGN capacity barely changes with different bandwidth size, and only a minor deviation is observed for the Kaiser window. Accordingly, the AWGN  rate-optimized SE  $\NeAWGN$ in~\eqref{eq:NeAWGN} corresponds to a nearly bandwidth-invariant fraction of the bandwidth. For fading channel with small frequency-selectivity, as in Fig.~\ref{fig:CapavsNe_Cos11_5DBsnr_30ns}, the capacity, and  hence  the optimal ratio $\NeCapa/\Nsc$, remains nearly bandwidth-invariant, similar to the PAPR optimization case. 
However, under stronger frequency selectivity (Figs.~\ref{fig:CapavsNe_Cos11_5DBsnr}-\ref{fig:CapavsNe_Cos11_0DBsnr}), the capacity decreases with increasing bandwidth, and the location of the optimal SE exhibits moderate fluctuations. 
When the bandwidth is small, $\NeCapa$ is close to $\NeAWGN$. As the bandwidth increases, the capacity degrades faster without SE than with 50\% SE, until a point beyond which the trend reverses. As a result, $\NeCapa$ temporarily increases, deviates from $\NeAWGN$, and then converges back toward it. We attribute this behavior to a trade-off between 
the frequency-diversity gain enabled by SE, which helps mitigate frequency-selective fading,
and,  
the faster convergence of $g_0$ toward its mean value as $\Nsc$ increases, since the summation $g_0$ involves a larger number of terms when $\Ne$ is smaller. 
Overall, $\NeCapa$ remains in the vicinity of $\NeAWGN$,  and the capacity curves are relatively flat around their maxima, implying that the capacity obtained with $\NeAWGN$ and   $\NeCapa$ are very similar.

Fig.~\ref{fig:NeCapOpt_vs_SNR_Cos11} shows $\NeCapa$ (in \% of $\Nsc$) versus SNR for a Hann FDSS window with  $\beta=-11$ dB under both AWGN  and   TDL-C(300ns)  channels, for $\Nsc= 96 $ and  $\Nsc = 600$. In all cases, the optimal SE size decreases with SNR and tends to toward zero, reflecting a fundamental trade-off between frequency-diversity gain and multiplexing efficiency.
Edge subcarriers in  the excess band experience larger FDSS attenuation 
and thus operate at lower effective SNR, where frequency diversity from SE is most beneficial. 
At low SNR, a large number of subcarriers  benefit from SE frequency diversity, whereas at high SNR,  
this diversity gain has less effect and a smaller SE size is preferred to preserve multiplexing gain.

\subsection{Simulation Evaluations of the Maximum Rate: FDSS-SE versus FDSS} 
\begin{figure}[t]
\vspace{-0.4cm}  
\includegraphics[width=.5\textwidth]{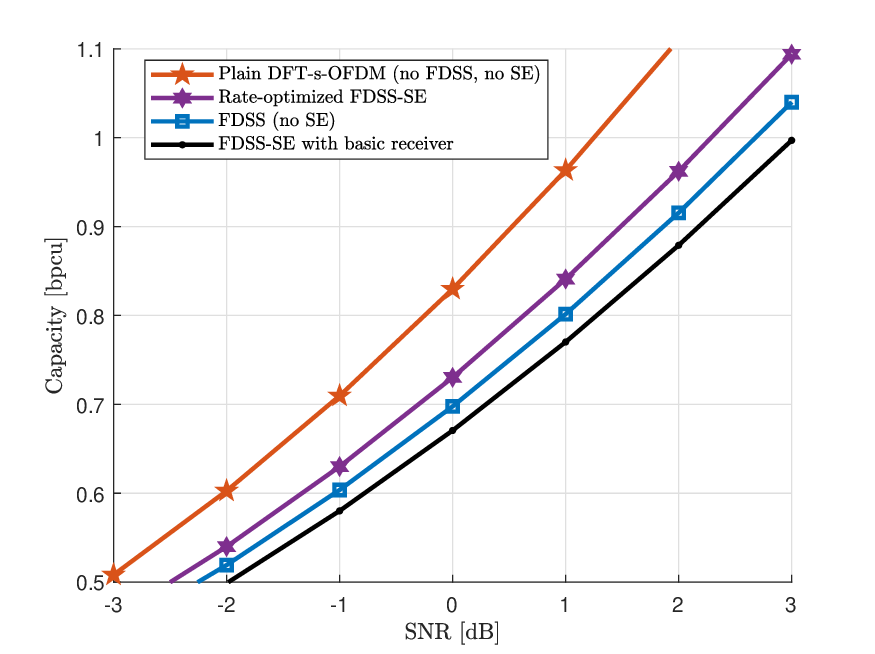}
\vspace{-0.7cm}
\caption{Capacity of plain DFT-s-OFDM, rate-optimized FDSS-SE, FDSS-SE with basic receiver, and FDSS only.  Hann FDSS window with $\beta=-11$ dB.	\label{fig:CapavsSNR}}
\vspace{-0.0cm}
\end{figure}

Fig.~\ref{fig:CapavsSNR} compares capacities as a function of SNR for an Hann FDSS window with $\beta=-11$ dB.  The channel model is TDL-C(300ns) with $\Nsc = 96$. Both FDSS or FDSS-SE  incur a capacity loss relative to plain DFT-s-OFDM that increases with SNR.  
At a spectral efficiency of 1 bpcu, the corresponding SNR losses relative to plain DFT-s-OFDM are 1, 1.4, and 1.8 dB for the rate-optimized FDSS-SE with MRC-MMSE receiver~\cite{Nokia21}, FDSS without SE, and FDSS-SE with basic receiver (see Rem.~\ref{RemBasicRx}), respectively.

\begin{figure}[t]
\vspace{-0.4cm}  
\includegraphics[width=.5\textwidth]{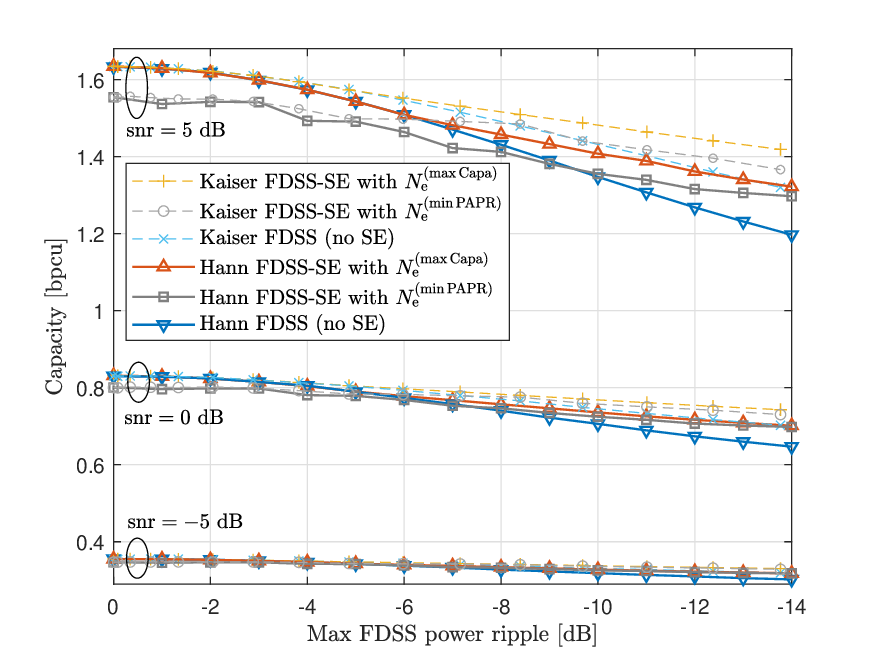}
\vspace{-0.7cm}
\caption{Comparison of achievable rate between  DFT-s-OFDM with optimized FDSS-SE and with FDSS only, as a function of FDSS shaping. 	\label{fig:CapavsShaping}}
\vspace{-0.0cm}
\end{figure}

Fig.~\ref{fig:CapavsShaping} shows the maximum achievable rate of FDSS-SE when using $\NeCapa$, compared to FDSS without SE, as a function of FDSS shaping. Both FDSS window types are considered, and  plain DFT-s-OFDM corresponds to an FDSS with 0dB attenuation. 

For low data rate operating at low SNR, FDSS attenuation has little impact on the maximum achievable rate, and the SE size also has minimal impact. 
For higher rates operating at higher SNR, the effect of FDSS shaping and SE size becomes more pronounced. At 5dB SNR, the spectral efficiency loss from FDSS Hann windowing with $-14$dB power ripple attenuation,  compared to plain DFT-s-OFDM without FDSS attenuation, is of $26\%$; however by using FDSS-SE with $\NeCapa$ for the same window, this loss can be reduced to $19\%$. 
The Kaiser windows outperforms the Hann window for strong shaping, which is becoming non-negligeable, especially in the higher-SNR regime.  At 5dB SNR, the spectral efficiency loss for the Kaiser window with similar $-14$dB power attenuation is of  $19\%$, which can be further reduced to $13\%$ with optimized SE size $\NeCapa$.

\begin{figure*}[t]
\centering
\vspace{-0.2cm} 
\subfigure[Optimum SE sizes, deformed Hann window \label{fig:NeCapOpt_vs_shaping_Hann}]{\includegraphics[width=.45\textwidth]{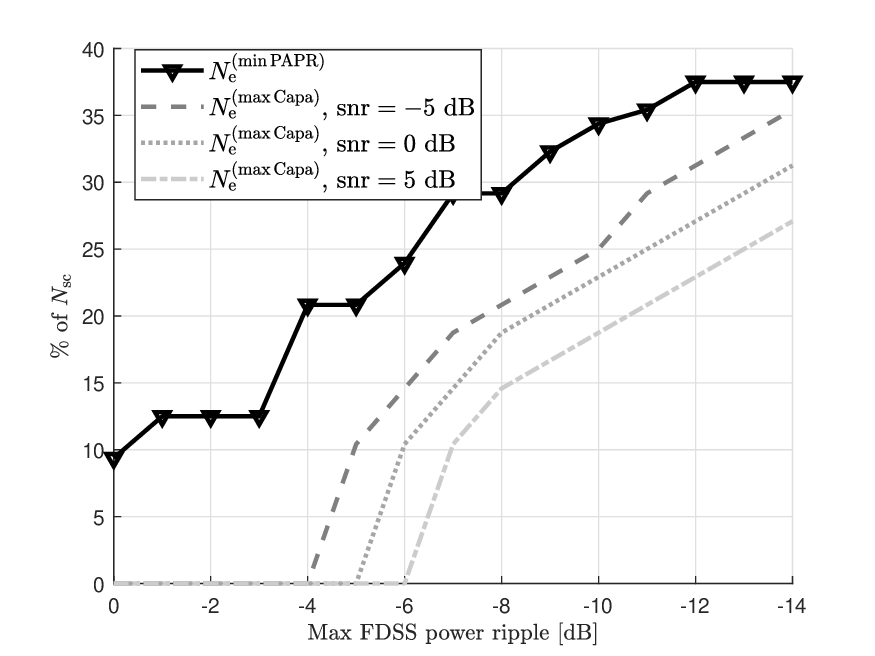}}
\subfigure[Optimum SE sizes, Kaiser window \label{fig:NeCapOpt_vs_shaping_Kaiser}]{\includegraphics[width=.45\textwidth]{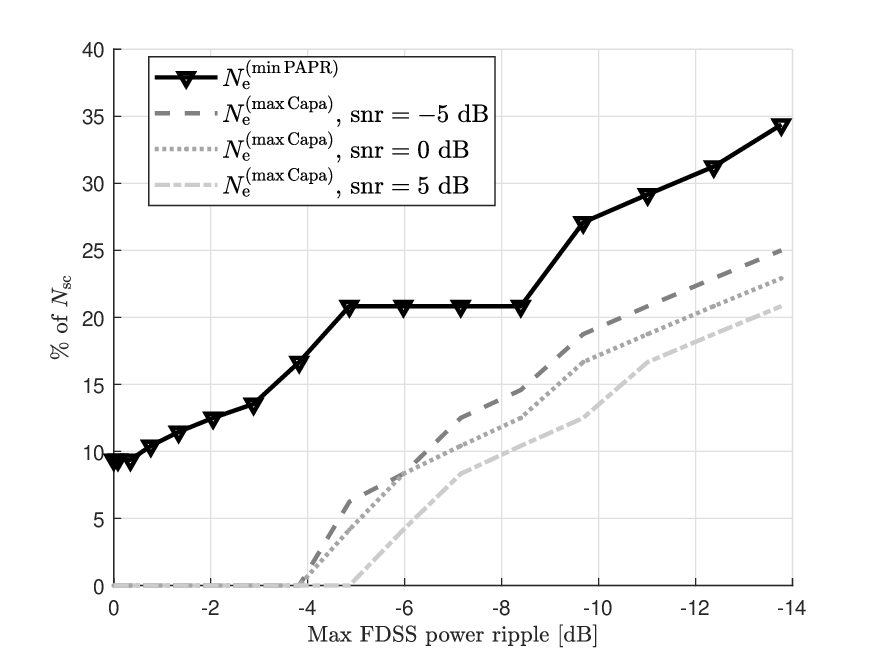}}
\subfigure[PAPR, deformed Hann window \label{fig:OptPAPR_vsPAPRNcapa_Hann}]{\includegraphics[width=.45\textwidth]{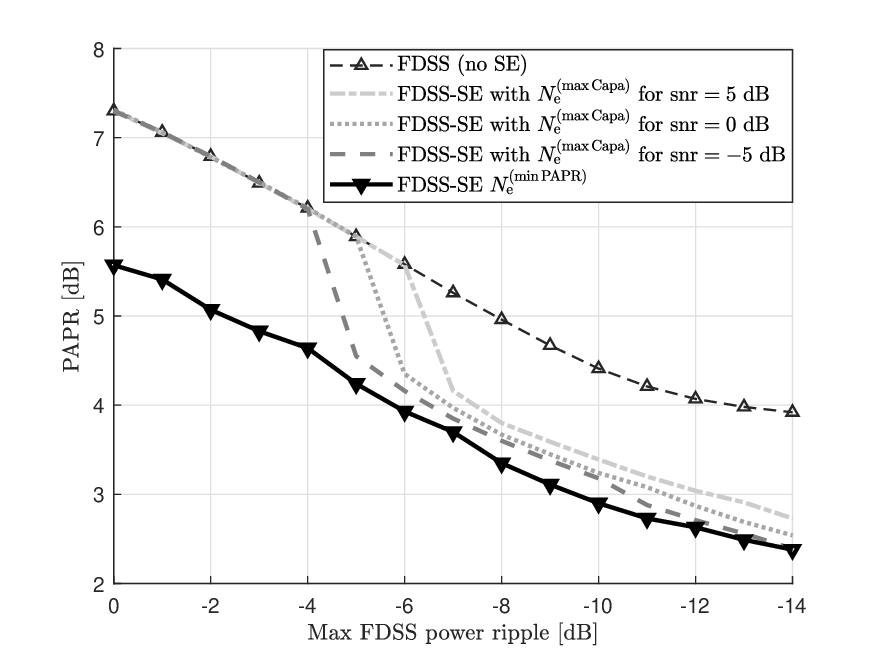}}
\subfigure[PAPR, Kaiser window \label{fig:OptPAPR_vsPAPRNcapa_Kaiser}]{\includegraphics[width=.45\textwidth]{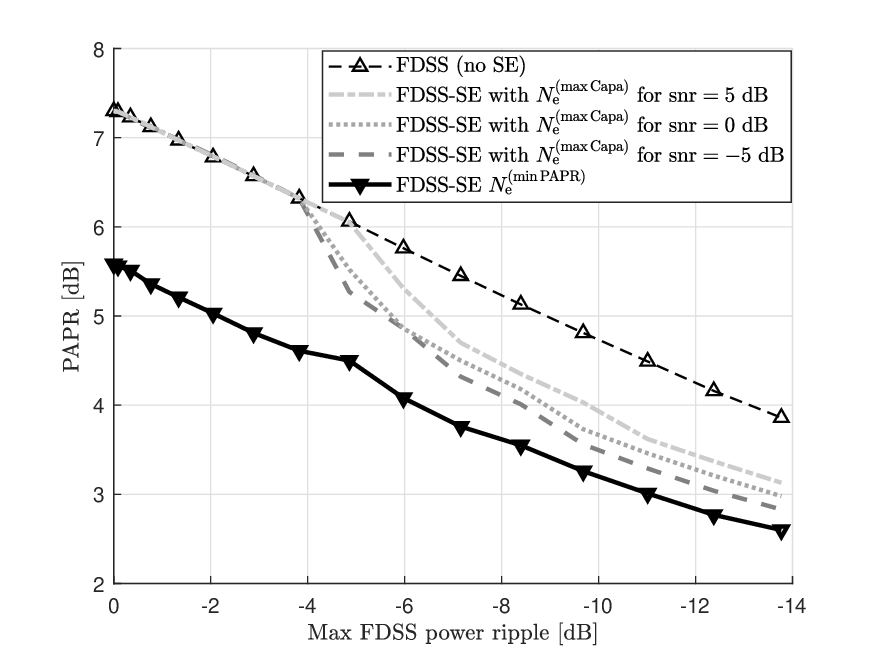}}
\vspace{-0.0cm}
	\caption{Comparison of $\NeCapa$ and $\NePAPR$, and the resulting PAPR trade-off as a function of FDSS shaping.   
	\label{fig:NeCapOpt_vs_shaping}
	}
\vspace{-0.2cm}
\end{figure*}

\subsection{Rate-PAPR Trade-Off} 
We showed that they are two distinct optima for SE size: one  for PAPR minimization ($\NePAPR$), and another for rate maximization ($\NeCapa$), which implies the existence of a range of SE sizes that can  simultaneously improve both PAPR and rate. Accordingly, $\NeCapa$ and $\NePAPR$ provides, respectively, a lower and upper bound on the SE size relevant for PAPR–rate trade-off. 

Fig.~\ref{fig:NeCapOpt_vs_shaping}  compares $\NePAPR$ and $\NeCapa$ as functions of the FDSS shaping for both window families with $\Nsc= 96$. 
For  $\NeCapa$, the TDL-C(300ns) channel is considered at SNR $-5$, 0, and 5 dB. 
The $\NePAPR$ curves in Figs.~\ref{fig:NeCapOpt_vs_shaping_Hann} and ~\ref{fig:NeCapOpt_vs_shaping_Kaiser} are obtained by  minimizing the bound~\eqref{eq:PAPRub} for QPSK.  
For both window types, the optimal SE size for PAPR minimization is larger than that for rate maximization, and the gap  widens with SNR. In general, the Kaiser window requires smaller SE sizes for both criteria.  
Interestingly, SE improves the rate of FDSS (i.e.  $\NeCapa>0$) only when the FDSS attenuation is sufficiently large.   
 
Fig.~\ref{fig:CapavsShaping}  reports the capacity if using $\NePAPR$ as identified in Figs.~\ref{fig:NeCapOpt_vs_shaping_Hann} and ~\ref{fig:NeCapOpt_vs_shaping_Kaiser}. 
With weak FDSS shaping, this choice is detrimental (e.g. a loss of $\approx 5$\% with no FDSS compared to plain DFT-s-OFDM at 5 dB SNR), whereas with stronger shaping, $\NePAPR$ can increase the rate. At $-14$dB attenuation and 5 dB SNR, the rate loss with Hann window is 21\% relative to plain DFT-s-OFDM (vs. 26\% with no SE, and  19\% with optimal SE), while for the Kaiser window it is of 17\% (vs. 19\% with no SE, and  13\% with optimal SE).

Similarly, although not optimal for PAPR minimization, the SE size $\NeCapa$  can still provide significant PAPR reduction, as shown on Figs.~\ref{fig:OptPAPR_vsPAPRNcapa_Hann} and \ref{fig:OptPAPR_vsPAPRNcapa_Kaiser},  which report the $10^{-3}$ CCDF PAPR values using the SE size from Figs.~\ref{fig:NeCapOpt_vs_shaping_Hann} and ~\ref{fig:NeCapOpt_vs_shaping_Kaiser}. 
 The PAPR reduction with $\NeCapa$ is nearly optimal for the Hann window at large attenuation, but weaker for the Kaiser window. 

A more specific SE size for rate-PAPR trade-off could in principle be obtained by incorporating an explicit SNR loss into the capacity formula to account for the required power back-off  associated with a given PAPR level.   As discussed in Sec. III, the CM has been used in past 3GPP studies  as a proxy for power back-off estimation, but its  linear fitting used in 3GPP  has not been updated since 4G, and  CM exhibits a non-monotonic behavior with FDSS shaping that is unlikely to yield reliable conclusions. Alternatively, linear fittings of PAPR on power de-rating are scarcely available and would require further validation in the presence of FDSS.  Moreover, a very precise optimum SE ratio is unlikely to be implementable in practice due do resources configuration constraints. In the meantime, the CM-based approach is gradually being abandoned in 3GPP in favor of more comprehensive but significantly less tractable RF power back-off simulations.

The analysis provided here helps delimit a practical range of SE sizes and provides useful rule of thumb for system design. 
For example, in the scenarios considered in Fig.~\ref{fig:NeCapOpt_vs_shaping} and  Fig.~\ref{fig:CapavsShaping}, when the SNR is low, the capacity variation is small and it is preferable to select a SE size close to $\NePAPR$, ranging from 10\% to 35\% depending of the FDSS window attenuation. As the SNR increases and for weak FDSS attenuation, the capacity penalty induced by SE becomes more pronounced, and smaller SE sizes on the order of 5\% may be preferable. Conversely, with strong FDSS attenuation, both $\NeCapa$ and $\NePAPR$ in the range of 20\%-35\% yield similar improvements in both rate and PAPR, such that further  optimization in between would provide only marginal gains. Finally, it is worth noting that the analysis confirms that a $25\%$ SE, advocated as a good compromise in~\cite{Nokia21} and in 3GPP Rel.-18 study, lies well within the SE range identified  in Fig.~\ref{fig:NeCapOpt_vs_shaping} for the corresponding scenario. 

\section{Conclusion}

This paper presented a systematic performance evaluation of FDSS with spectrum extension (FDSS-SE) by  employing single-parameter FDSS window families, encompassing various SE methods in the literature.  For $\pi/2$-BPSK, the symmetrical double-side SE is optimal, whereas for higher-order QAM, the single-side SE performs better while optimality is achieved by another asymmetric SE. 
More importantly, we showed that there exist two distinct optima for the SE size: one  for PAPR minimization and another for rate maximization, demonstrating the possibility of simultaneously improving both metrics.    
The optimum SE for PAPR performance depends mainly of the FDSS window and is largely invariant to bandwidth and constellation order for regular QAM. The optimum SE for rate maximization depends on the FDSS window and the SNR, while remaining also nearly invariant to bandwidth. This suggests that if SE is introduced on top of FDSS in 6G, it should be adaptively configured based on the user's FDSS implementation, if proprietary, and the transmission link quality.

\appendix
\subsection{3-tap Filters as Deformed Hann Window} %
\label{Proof:3tapFilter}
Consider a 3-tap filter defined as $ f= [-b,\ 1,\ -b]$. The corresponding un-normalized FDSS window is
\begin{eqnarray}
W_{\rm 3-tap}[k'] &=& \sum_{m=0}^2 f[m] e^{-\jrm \frac{2 \pi}{\Nsc}k'm}\\
&=&e^{-\jrm \frac{2 \pi}{\Nsc}k'} \left(1-2b \cos  \frac{2 \pi}{\Nsc}k' \right). 
\end{eqnarray}
The linear phase ramp $e^{-\jrm \frac{2 \pi}{\Nsc}k'}$ has only the effect of a circularly shift on the OFDM symbol so it has no impact on the PAPR and can be ignored. 
Then, using the change of variable, $2b = \frac{1-\beta}{1+\beta}$, we have a parametrization that quantifies directly the maximum ripple of the window as then  $\frac{\min_k  W_\beta[k']}{\max_k  W_\beta[k']} = \beta$. 

We get~\eqref{eq:ModHanWin} by applying an half index shift $k=k'+1/2$ for symmetry such that $W_\beta[k]=W_\beta[\Nsc-1-k]$; and a normalization  following from from the geometric sum formula showing that for a constant $\alpha$ we have 
\begin{equation}
\sum_{k=0}^{\Nsc-1} \left(1-\alpha \cos  \frac{2 \pi}{\Nsc}k \right)^2 = \Nsc\left(1+\frac{\alpha^2}{2}\right).
\end{equation}
 
\subsection{PAPR Bounds} %
\label{App:PAPRbounds}
Noticing that that the average power of an OFDM symbol is $\Ebb{|s[n]|^2} = \frac{\Nsc}{\Nfft} $, the PAPR  simplifies
 to 
\begin{equation}  \label{eq:PAPRsimp}
{\rm PAPR } =  \frac{\Nfft}{\Nsc} \max_{0\leq n <\Nfft} |s[n]|^2.
\end{equation}  

Upper bound on $\max_{0\leq n <\Nfft} |s[n]|^2$ then follows by adapting~\cite[Th. 1]{KimTVT18} and~\cite[Th. 1]{TENCON18} provided for $\phi$-BPSK and QPSK, respectively, to any constellation and shifts. The case considered therein corresponds to $L =0$, with a shift $L$ then $ \angle \frac{p_i [n]}{p_j[n]} = -(i-j)2\pi\frac{L}{\Ndata}  +\angle \frac{p_0 \left[n-\frac{\Nfft}{\Ndata} i\right] }{p_0\left[n-\frac{\Nfft}{\Ndata} j \right]  } $ where the last term is evaluated in~\cite[Lem. 1]{KimTVT18}.  

The second general upper bound follows directly from the fact that $u_{i,j}^{\Ccal}\leq 1$.

\subsection{Proof of~\eqref{eq:pulsePhdiff} } %
\label{App:EqPhase}
From~\eqref{eq:pm}, the phase difference between two neighboring pulses with indices $m$ and $(m+1)$ is
\begin{equation}
\angle \frac{p_{m+1}[n]}{p_{m}[n]} = - \frac{2\pi L}{\Ndata} +\angle \frac{p_{0}\left[n -\frac{\Nfft}{\Ndata}(m+1)\right]}{p_{0}\left[n -\frac{\Nfft}{\Ndata}m\right]}.
\end{equation}
	Then, as shown in~\cite[Lem. 1]{KimTVT18} if the FDSS window is real and symmetric we have
	\begin{equation}
	\angle \frac{p_{0}\left[n -\frac{\Nfft}{\Ndata}(m+1)\right]}{p_{0}\left[n -\frac{\Nfft}{\Ndata}m\right]} = -\frac{(\Nsc-1)}{\Ndata} \pi + \{ 0 \text{ or } \pi \}
	\end{equation}
which can also be directly verified to hold true for the Dirichlet pulse~\eqref{eq:p0sinc}.
Therefore we get the pulse phase difference 
\begin{eqnarray} \label{}
\angle \frac{p_{m+1}[n]}{p_{m}[n]} &=& - \frac{2\pi}{\Ndata}\left(L +\frac{(\Nsc-1)}{2} \right) \; ({\rm mod}\,  \pi) \nonumber\\
 &=& - \frac{2\pi}{\Ndata}\left(L +\frac{(\Ne-1)}{2} \right) \; ({\rm mod}\,  \pi).
\end{eqnarray} 
\subsection{Proofs of  Lem.~\ref{Prop1}} %
\label{Proof:Prop1}

Expanding~\eqref{eq:r[m]} and using a change of variable, we have 
 \begin{eqnarray} \label{eq:r[k]2}
r[m]&=& \frac{1}{\sqrt{\Ndata}} \sum_{k=0}^{\Ndata-1} \tilde{R}_{\rm eq}[k-L] e^{\jrm \frac{2\pi}{\Ndata}km} \\
&=& \frac{e^{\jrm \frac{2\pi}{\Ndata}mL}}{\sqrt{\Ndata}} \sum_{k=0}^{\Ndata-1} \tilde{R}_{\rm eq}[k] e^{\jrm \frac{2\pi}{\Ndata}km} . \label{eq:r[k]3}
\end{eqnarray}
 Expanding \eqref{eq:Req[k]}  we get
 \begin{equation} \label{eq:Req[k]Decompose}
\tilde{R}_{\rm eq}[k]= \frac{G[k]}{G[k]+1} X[k+L] +\frac{\tilde{Z}[k]}{G[k]+1}.
\end{equation}
Plugging~\eqref{eq:Xdft} in~\eqref{eq:Req[k]Decompose}, and~\eqref{eq:Req[k]Decompose} in~\eqref{eq:r[k]3}, we can rewrite ~\eqref{eq:r[k]2} as 
 \begin{eqnarray} 
r[m] &=&     \sum_{n=0}^{\Ndata-1} g_{m-n} x[n] +n[m] \nonumber\\
&=& g_{0}x[m]+{\rm ICI}[m] +n[m]
\end{eqnarray}
where $g_{0}$ is the desired channel component, independent of $m$,  ${\rm ICI}[m] = \sum_{\substack{n=0\\n\neq m}}^{\Ndata-1} g_{m-n} x[n] $ with 
 \begin{equation} \label{eq:ga} 
g_{a}=     \frac{e^{\jrm\frac{ 2\pi }{\Ndata}L a}}{\Ndata}\sum_{k=0}^{\Ndata-1} \frac{G[k]}{G[k]+1} e^{\jrm 2\pi \frac{k a}{\Ndata}}, 
\end{equation}
and the processed noise is  
\begin{equation} \label{eq:n[m]}
n[m] =  \frac{ e^{\jrm\frac{2\pi}{\Ndata}L m}}{\sqrt{\Ndata}}
\sum_{k=0}^{\Ndata-1} \frac{\tilde{Z}[k]}{G[k]+1} e^{\jrm 2\pi \frac{k m}{\Ndata}}.
\end{equation}

The desired signal power is directly given by  $g_{0}^2$. Noticing that the equalized noise power is $\sigma_{\tilde{Z}}^2 = G[k] \sigma_{Z}^2$, 
the power of the processed noise~\eqref{eq:n[m]} becomes $\sigma_n^2 =  \frac{1}{\Ndata} \sum_{k=0}^{\Ndata-1}\frac{ G[k] }{(G[k]+1)^2}$.

 The noise-plus-interference power term can be computed as similarly done in~\cite{Nisar2007} and simplified to $(g_{0}-g_{0}^2)$. To verify this, the ICI power can be first simplified as 
\begin{eqnarray}
\sigma_{\rm ICI}^2 &=& \sum_{\substack{n=0\\n\neq m}}^{\Ndata-1} |g_{m-n}|^2 = \left(\sum_{n=0}^{\Ndata-1} |g_{m-n}|^2\right) -  g_{0}^2 \nonumber\\
&=&  \frac{1}{\Ndata} \sum_{k=0}^{\Ndata-1}\frac{G[k]^2}{(G[k]+1)^2} -  g_{0}^2
\end{eqnarray}
where the last equality follows by direct expansion of~\eqref{eq:ga} and simplification by the exponential sum formula which gives  $\sum_{n = 0}^{\Ndata-1}e^{\jrm 2\pi \frac{n a}{\Ndata}} = 0 $ for any non-zero integer $a$.
Then by direct combination we have 
\begin{eqnarray}
\sigma_{\rm ICI}^2  + \sigma_n^2 &=& 
\frac{1}{\Ndata} \sum_{k=0}^{\Ndata-1}\frac{G[k]^2+ G[k] }{(G[k]+1)^2}
-  g_{0}^2  \nonumber\\ 
&=& \frac{1}{\Ndata} \sum_{k=0}^{\Ndata-1}\frac{G[k] }{G[k]+1}
-  g_{0}^2 \nonumber\\ 
&=&g_{0}-g_{0}^2.
\end{eqnarray}
 The corresponding SINR~\eqref{eq:SNReff} follows accordingly. 

Using this SINR expression, which is independent of $m$, the achievable rate~\eqref{eq:Capa} is obtained as the sum-capacity of the DFT-s-OFDM subcarriers. This follows by treating the subcarriers as parallel channels, and bounding the mutual information of each to the one of a Gaussian additive noise channel with known noise variance, i.e. conditioned also on the knowledge of the effective interference channel gains. The rate is thus an
achievable rate under the assumption of perfect knowledge of
channel state information at the receiver.

\addcontentsline{toc}{chapter}{Bibliography}
\bibliographystyle{IEEEtran} 
\bibliography{mybibfile}

@misc{3GPPTS38.101-2,
author={ {\relax 3GPP TS 38.101-2 v. 19.0.0 }},
title={User Equipment ({UE}) Radio Transmission and Reception; Part 2: Range 2 Standalone, Technical Specification Group Radio Access
Network},
year = 2025,
month ={Mar.}
}

@misc{3GPPTS25.101,
author={ {\relax 3GPP TS 25.101 v. 17.0.0}},
title={Universal Mobile Telecommunications System ({UMTS}); User Equipment ( {UE}) radio transmission and reception ({FDD})},
year = 2022,
month ={Apr.}
}

@misc{ReportRAN1_102,
author={ {\relax 3GPP TSG RAN WG1}},
title={Final report},
howpublished = {Meeting 102-e, document R1-2007501},
year = {2020},
month ={Aug.}
}

@misc{RP-213579,
author={ {\relax 3GPP TSG RAN WG1 --  China Telecom}},
title={{\relax New WI: Further NR coverage enhancements}},
howpublished = {Meeting  94e, document RP-213579},
year = {2021},
month ={Dec.}
}

@misc{R1-060023,
author={{\relax 3GPP TSG RAN WG1 -- Motorola} },
title={Cubic Metric in {3GPP-LTE}},
howpublished = {Meeting  LTE Adhoc, document R1-060023},
year = 2006,
month ={Jan.}
}

@misc{R1-040642, 
author={{\relax 3GPP TSG RAN WG1 -- Motorola} },
title={Comparison of {PAR} and Cubic Metric for Power De-rating},
howpublished = {Meeting  37, document R1-040642},
year = 2004,
month ={May}
}

@misc{R1-163437,
author={{\relax 3GPP TSG RAN WG1 -- Ericsson, Nokia, Lenovo, LG, ZTE} },
title={{WF} on {DMRS} multi-tone evaluation methods},
howpublished = {Meeting  84-bis, document R1-163437},
year = 2016,
month ={Apr.}
}

@misc{R1-1901117,
author={{\relax 3GPP TSG RAN WG1 -- Ericsson} },
title={Additional simulation results on low {PAPR RS}},
howpublished = {Meeting  Ad-Hoc Meeting 1901, document R1-1901117},
year = 2019,
month ={Jan.}
}

@misc{R1-2210880,
author={{\relax 3GPP TSG RAN WG1 -- Huawei, HiSilicon} },
title={Discussion on coverage enhancement in power domain},
howpublished = {Meeting  Ad-Hoc Meeting 111, document R1-2210880},
year = 2022,
month ={Nov.}
}

@misc{R1-2208412,
author={{\relax 3GPP TSG RAN WG1 -- Huawei, HiSilicon} },
title={Discussion on coverage enhancement in power domain},
howpublished = {Meeting  Ad-Hoc Meeting 110bis-e, document R1-2208412},
year = 2022,
month ={Oct.}
}

@misc{R1-2300090,
author={{\relax 3GPP TSG RAN WG1 -- Huawei, HiSilicon} },
title={Discussion on coverage enhancement in power domain},
howpublished = {Meeting  113, document R1-2300090},
year = 2023,
month ={May}
}

@misc{R1-2305484,
author={{\relax 3GPP TSG RAN WG1 -- Ericsson} },
title={Power Domain Enhancement Schemes and Performance},
howpublished = {Meeting  113, document R1-2305484},
year = 2023,
month ={May}
}

@misc{RP-251862,
author={{\relax 3GPP TSG RAN} },
title={New {WID}: Coverage enhancements for {NR} Phase 3},
howpublished = {Meeting  108, document RP-251862},
year = 2025,
month ={June}
}

@misc{RP-251881,
author={{\relax 3GPP TSG RAN} },
title={New {SID}: Study on {6G} Radio},
howpublished = {Meeting  108, document RP-251881},
year = 2025,
month ={June}
}

@misc{R1-050702,
author={ {\relax 3GPP TSG RAN WG1 -- NTT DoCoMo, NEC, SHARP}},
title={{\relax DFT-Spread OFDM with Pulse Shaping Filter in Frequency Domain in Evolved UTRA Uplink}},
howpublished = {Meeting  42, document R1-050702},
year = {2005},
month ={Sep.}
}

@misc{6GWS-250004,
author={{\relax 3GPP TSG RAN -- Nokia} },
title={{6G Radio and RAN}},
howpublished = {3GPP workshop on 6G, document 6GWS-250004},
year = 2025,
month ={Mar.}
}

@misc{6GWS-250036,
author={{\relax 3GPP TSG RAN -- Samsung} },
title={Vision and Technologies for {6G} Radio},
howpublished = {3GPP workshop on 6G, document 6GWS-250036},
year = 2025,
month ={Mar.}
}

@INPROCEEDINGS{Mauritz06,
  author={Mauritz, Oskar and Popovic, Branislav M.},
  booktitle={Proc. IEEE Veh. Tech. Conference}, 
  title={Optimum Family of Spectrum-Shaping Functions for {PAPR} Reduction of {DFT}-Spread {OFDM} Signals}, 
  year={2006},
  volume={},
  number={},
  pages={1-5}}

@INPROCEEDINGS{Kawamura06,
  author={Kawamura, Teruo and Kishiyama, Yoshihisa and Higuchi, Kenichi and Sawahashi, Mamoru},
  booktitle={Proc. Int. Symp. Wireless Commun. Syst.}, 
  title={Investigations on Optimum Roll-off Factor for {DFT}-Spread {OFDM} Based {SC-FDMA} Radio Access in Evolved {UTRA} Uplink}, 
  year={2006},
  volume={},
  number={},
  pages={383-387}}

@ARTICLE{Nokia21,
  author={Nasarre, Ismael Peruga and Levanen, Toni and Pajukoski, Kari and Lehti, Arto and Tiirola, Esa and Valkama, Mikko},
  journal={IEEE Open J. Commun. Soc.}, 
  title={Enhanced Uplink Coverage for {5G NR}: Frequency-Domain Spectral Shaping With Spectral Extension}, 
  year={2021},
  volume={2},
  number={},
  pages={1188-1204}}

@ARTICLE{SlimaneTVT07,
  author={S. B. Slimane},
  journal={IEEE Trans.  Veh. Tech.}, 
  title={Reducing the Peak-to-Average Power Ratio of {OFDM} Signals Through Precoding}, 
  year={2007},
	month ={Mar.},
  volume={56},
  number={2},
  pages={686-695}}

@ARTICLE{KimTVT18,
  author={J. Kim and Y. H. Yun and C. Kim and J. H. Cho},
  journal={IEEE Trans.  Veh. Tech.}, 
  title={Minimization of {PAPR} for {DFT}-Spread  {OFDM} With {BPSK} Symbols}, 
  year={2018},
	month ={Dec.},
  volume={67},
  number={12},
  pages={11746-11758}}

@Article{Qin2022,
  author   = {Qin, Yi and Pitaval, Renaud-Alexandre},
  journal  = {IEEE Commun. Lett.},
  title    = {Low Cubic Metric {Reed-Muller} Sequence Design for Pilot-Less Transmission},
  year     = {2022},
  month    = feb,
  number   = {2},
  pages    = {364--368},
  volume   = {26},
}

@INPROCEEDINGS{TENCON18,
  author={Kim, Jubum and Yun, Yeo Hun and Kim, Chanhong and Cho, Joon Ho},
  booktitle={Proc. IEEE Region 10 Int. Conf. TENCON}, 
  title={{PAPR} Reduction by Constellation Rotation and Pulse Shaping for {DFT-Spread OFDM with QPSK} Symbols}, 
  year={2018},
  volume={},
  number={},
  pages={0090-0095}}

@INPROCEEDINGS{SamsungSPAWC2024,
  author={F. Carpi and S. Rostami and J. Cho and S. Garg and E. Erkip and C. J. Zhang},
  booktitle={IEEE Int. Workshop Sig. Proc. Advances Wireless Commun. (SPAWC)}, 
  title={Learned Pulse Shaping Design for {PAPR} Reduction in {DFT-s-OFDM}}, 
  year={2024},
  volume={},
  number={},
  pages={406-410}}

@INPROCEEDINGS{Nisar2007,
  author={Nisar, Muhammad Danish and Nottensteiner, Hans and Hindelang, Thomas},
  booktitle={Proc. IST Mobile Wireless Commun. Summit}, 
  title={On Performance Limits of {DFT} Spread {OFDM} Systems}, 
  year={2007},
  volume={},
  number={},
  pages={1-4}}

@ARTICLE{PAPROverview,
  author={Jiang, Tao and Wu, Yiyan},
  journal={IEEE Trans. Broadcasting}, 
  title={An Overview: Peak-to-Average Power Ratio Reduction Techniques for {OFDM} Signals}, 
  year={2008},
  volume={54},
  number={2},
  pages={257-268}}

@ARTICLE{SlepianPollak,
  author={D. Slepian and H. Pollak},
  journal={Bell Sys. Tech. Journal}, 
  title={Prolate-spheroidal wave functions, {F}ourier analysis and uncertainty--i}, 
  year={1961},
  volume={40},
  pages={ 34-64}}

@ARTICLE{IEICE25,
  author={Kawasaki, Hikaru and Kojima, Fumihide and Matsumura, Takeshi},
  journal={IEICE Trans.  Commun.}, 
  title={Computation-free phase rotation by shifted circular repetition for DFT-precoding scheme to peak power reduction of {OFDM} signals}, 
  year={2025},
  volume={},
  number={},
  pages={1-17}}

@ARTICLE{IEICE25_NTT,
  author={Liu, Juan and Hou, Xiaolin and Liu, Wenjia and Li, Chen and Chen, Lan},
  journal={IEICE Trans.  Commun.}, 
  title={Low-{PAPR} {6G} waveform and modulation design with {5G NR} backward compatibility}, 
  year={2025},
  volume={},
  number={},
  pages={1-15}}

@misc{R1-2506306,
author={{\relax 3GPP TSG RAN WG1 -- NTT DOCOMO} },
title={Discussion on Waveform},
howpublished = {Meeting  122, document R1-2506306},
year = 2025,
month ={Aug.}
}

@INPROCEEDINGS{ICCT17,
  author={Tang, Bo and Zhang, Xiangyin and Qin, Kaiyu and Wu, Shaowei},
  booktitle={Proc. IEEE Int. Conf. Commun. Tech.}, 
  title={Experimental comparison of various metrics on power de-rating prediction}, 
  year={2017},
  volume={},
  number={},
  pages={1219-1222}}

@INPROCEEDINGS{LiWei,
  author={Li, Wei and Wang, Jingqi and He, Lintong and Yao, Zhiming},
  booktitle={IEEE United Conference on Millimeter Waves and Terahertz Technologies (UCMMT)}, 
  title={{PAPR} Reduction for {OFDM-IM}: An Enhanced Active Constellation Extension Method}, 
  year={2025},
  volume={1},
  number={},
  pages={1-3}}

@ARTICLE{HuangPengfei,
  author={Huang, Pengfei and Li, Qiang and Wen, Miaowen and Dang, Xiaoyu and Huang, Dong},
  journal={IEEE Commun. Let.}, 
  title={Deep-Learning-Based   {PAPR} Reduction for Pilot-Embedded {AFDM} Signals}, 
  year={2025}}

@ARTICLE{AliAfan,
  author={Ali, Afan and Arous, Abdelali and Arslan, Hüseyin},
  journal={IEEE Trans.  Green Commun. Networking}, 
  title={Spreading the Wave: Low-Complexity {PAPR} Reduction for {AFDM} and {OCDM} in {6G} Networks}, 
  year={2025}}

@ARTICLE{Kant,
  author={Kant, Shashi and Bengtsson, Mats and Fodor, Gabor and Göransson, Bo and Fischione, Carlo},
  journal={IEEE Trans.   Wireless Commun.}, 
  title={{EVM} Mitigation With {PAPR} and {ACLR} Constraints in Large-Scale {MIMO-OFDM} Using {TOP-ADMM}}, 
  year={2022},
  volume={21},
  number={11},
  pages={9460-9481}}

\end{document}